\documentclass[10pt,journal,compsoc]{IEEEtran}

\ifCLASSOPTIONcompsoc
  \usepackage[nocompress]{cite}
\else
  \usepackage{cite}
\fi

\ifCLASSINFOpdf
  \usepackage[pdftex]{graphicx}
  \usepackage[bookmarks=false]{hyperref}
\else
  \usepackage[dvips]{graphicx}
\fi

\usepackage[justification=centering]{caption}

\ifCLASSOPTIONcompsoc
  \usepackage[caption=false,font=normalsize,labelfont=sf,textfont=sf]{subfig}
\else
  \usepackage[caption=false,font=footnotesize]{subfig}
\fi
\usepackage[usenames,dvipsnames,svgnames]{xcolor}

\newif\ifcommentson
\commentsontrue

\newif\ifextended
\newif\ifshortver

\shortvertrue
\extendedtrue

\newcommand{\optional}[1]{\ignorespaces}

\begin{document}
\bstctlcite{IEEEexample:BSTcontrol}

%
\title{%
On the Fly Orchestration of Unikernels:\\%
Tuning and Performance Evaluation\\%
of Virtual Infrastructure Managers%
}

\author{%
\vspace{-1.5ex}
\IEEEauthorblockN{
Pier Luigi Ventre\IEEEauthorrefmark{1}\IEEEauthorrefmark{2}\quad
Paolo Lungaroni\IEEEauthorrefmark{2}\quad
Giuseppe Siracusano\IEEEauthorrefmark{1}\IEEEauthorrefmark{4}\quad
Claudio Pisa\IEEEauthorrefmark{2}\quad\\
Florian Schmidt\IEEEauthorrefmark{4}\quad
Francesco Lombardo\IEEEauthorrefmark{2}\quad
Stefano Salsano\IEEEauthorrefmark{1}\IEEEauthorrefmark{2}\quad
}

\IEEEauthorblockA{
\IEEEauthorrefmark{1}University of Rome Tor Vergata, Italy\quad
\IEEEauthorrefmark{2}CNIT, Italy\quad
\IEEEauthorrefmark{4}NEC Labs Europe, Germany\quad
}}


%

\markboth{\textbf{Submitted to IEEE Transactions on Cloud Computing}}%
{Ventre \MakeLowercase{\textit{et al.}}: On the Fly Orchestration of Unikernels: Tuning and Performance Evaluation of Virtual Infrastructure Managers}


\IEEEtitleabstractindextext{
\vspace{-2ex}
\begin{abstract}
Network operators are facing significant challenges meeting the demand for more bandwidth, agile infrastructures, innovative services, while keeping costs low. Network Functions Virtualization (NFV) and Cloud Computing are emerging as key trends of 5G network architectures, providing flexibility, fast instantiation times, support of Commercial Off The Shelf hardware and significant cost savings. NFV leverages Cloud Computing principles to move the data-plane network functions from expensive, closed and proprietary hardware to the so-called Virtual Network Functions (VNFs). In this paper we deal with the management of virtual computing resources (Unikernels) for the execution of VNFs. This functionality is performed by the Virtual Infrastructure Manager (VIM) in the NFV MANagement and Orchestration (MANO) reference architecture. We discuss the instantiation process of virtual resources and propose a generic reference model, starting from the analysis of three open source VIMs, namely OpenStack, Nomad and OpenVIM. We improve the aforementioned VIMs introducing the support for special-purpose Unikernels and aiming at reducing the duration of the instantiation process. We evaluate some performance aspects of the VIMs, considering both stock and tuned versions. The VIM extensions and performance evaluation tools are available under a liberal open source licence.
\end{abstract}
\vspace{-1ex}
\begin{IEEEkeywords}
Network Function Virtualization, Open Source, Virtual Infrastructure Manager, Virtual Network Function, Performance, Tuning, OpenStack, Nomad, OpenVIM, ClickOS, Unikernel
\end{IEEEkeywords}
}

\maketitle

\IEEEdisplaynontitleabstractindextext
%
\IEEEpeerreviewmaketitle

\IEEEraisesectionheading{%
\vspace{-20ex}
\section{Introduction}\label{sec:introduction}%
}

\IEEEPARstart{N}{etwork} Function Virtualization (NFV) is a new pa\-ra\-digm~\cite{nfv}, drastically changing the design of the current telecommunication networks. NFV introduces software components in place of specialized network hardware. These software modules, called Virtual Network Functions (VNFs), \optional{execute the same functions of network appliances and}run using virtual computing resources. Telco operators are looking to Cloud Computing~\cite{cloud} best practices and principles to build ``clouds'' in their networks (core, aggregation, edge). \optional{In this scenario, NFV would enable building an infrastructure from commodity blocks, able to rapidly deploy and dynamically scale services.} Flexibility, agility, fast instantiation times, consolidation, Commercial Off The Shelf (COTS) hardware support and significant cost savings are fundamental for meeting the requirements and facing the challenges of the new generation of telco networks. %
\optional{NFV together with Software Defined Networking (SDN)~\cite{sdn} can be seen as part of an even wider trend towards the softwarization of networks~\cite{galis}~\cite{kind}, which implies a complete rethinking of how Service Provider networks are now structured.}%
\optional{Their promise is the achievement of unmatched infrastructure scalability, flexibility and efficiency of networks (reducing equipment and operational costs) through the ubiquitous employment of software-based automation on COTS hardware.}%

\optional{Virtualization is the key technology behind Cloud Computing.} \optional{NFV paradigm, relying upon virtualization techniques, aims at enjoy same benefits.} Different virtualization approaches can be used to support VNFs: Virtual Machines (VMs), tinified VMs, Unikernels and containers (see~\cite{huicicomparison,manco2017my,shetty2017empirical} for a description of these concepts and an exhaustive comparison). Considering the performance aspects, containers are currently the preferred solution, as they deliver a lighter-weight virtualization technology and better run time performance in terms of throughput and delay compared to VMs. The Unikernels are specialized VMs and can provide similar performance figures than containers~\cite{huicicomparison}, but much better security properties.  
Unikernels have the ability to provide efficient, high-specialized, reusable, micro-operations, which can be very useful in NFV scenarios. They offer very good run time performance and very low memory footprint and instantiation time. The most important reason to choose the Unikernels as virtualization technology is that they have very good isolation and security properties, making them suitable to multi-tenancy scenarios. Their applicability goes beyond the NFV use cases, they can be used also to run ``high level applications'' like a Web Server. In \cite{xenwh}, Unikernels role is re-shaped and they are placed at the ground of the next generation Clouds. In \cite{cormack} Unikernels are mentioned as a key technology for the evolution of the modern operating systems. A witness of the increasing importance of Unikernels is the Unicore/Unikraft project \cite{unikraft} recently incubated by the Xen project. For completeness of this overview of virtualization technology, we mention the tinified VMs, such as the ones yielded by the Tinyx/LightVM approach~\cite{manco2017my}. They deliver a high degree of flexibility with good run time performance compared to full-fledged VMs, but their performance does not quite reach that of containers or Unikernels.

Current NFV technology mostly targets scenarios where VNFs are run by full-fledged Virtual Machines (VMs) (like for example in \cite{clearwater}) \optional{they are spawn to perform a specific task, and} and where the life-cycle operations, such as VNF creation, are in the time scale of minutes or several seconds. In this context, Network Services are realized by chaining and composing these \textit{big} VMs. Other solutions (for example \cite{cord}) leverage container-based solutions instead of full-fledged VMs. The concept of a \textit{Superfluid} NFV approach has been first illustrated in ~\cite{manco2014towards, manco2015case}. It aims to support highly dynamic scenarios, in which the VNFs are instantiated ``on the fly'' following the service requests, reducing the time scale of the life-cycle operations of virtualized resources down to few milliseconds.  %
\optional{They provide an infrastructure able to instantiate services on-the-fly and move the running software efficiently to different locations. This Cloud infrastructure spreads in different sections of the Service Provider network and yields to a massive consolidation in the hardware servers.}%
In these scenarios, VNFs tend to become small and highly specialized, i.e., elementary and reusable components targeting micro-operations. Complex services can be built through the ``chaining'' of these special-purpose VNFs. The instantiation times of the virtualized resources could be required to stay in the order of milliseconds. The architecture proposed in ~\cite{SF-architecture} follows the Superfluid NFV approach considering the decomposition of services into smaller reusable components and the highly dynamic instantiation of virtualized resources. The Unikernel virtualization technology is supported as a mean to achieve the performance targets (instantiation time, small memory footprints) and provide high security by proper isolation of the virtualized resources.

In this work we focus on ClickOS~\cite{martins2014clickos}, a Xen-based Unikernel tailored for NFV appliances, which is able to provide highly efficient raw packet processing. It has a small footprint (around 5 MB when running), can be instantiated within around 30 milliseconds, processes up to 10Gb/s of traffic and does not need a disk to operate. In addition, it benefits from the isolation provided by the Xen hypervisor and the flexibility offered by the Click modular router. Recent work \cite{manco2017my} demonstrates that it is possible to guarantee lower latency instantiation times (in the order of few msecs) for Unikernels, by using a customized hypervisor \optional{, like LightVM does,} and redesigning the toolstack to increase the efficiency.

Since Unikernels and a redesigned toolstack can provide an efficient implementation of the Network Functions Virtualization Infrastructure (NFVI), we are stimulated in shedding some light on the performances guaranteed by the NFV framework as a whole, considering Superfluid NFV use cases. In particular, we are interested on the instantiation process of Unikernels in the NFV framework. Borrowing the ETSI NFV terminology (see Figure~\ref{fig:nfv}) we focus on the Virtual Infrastructure Manager (VIM), which controls and manages the virtual resources in the NFVI. \optional{Taking Superfluid Cloud requirements as main driver, the scientific and technological question ``What are the performances of the VIMs in the NFV framework'' is definitely still open and needed a response.} Existing open source solutions, designed for less specialized Cloud infrastructures, provide a solid base to build VIMs for NFVI. However, a deeper analysis (see Section \ref{sec:vims} and Section \ref{sec:relatedwork}) reveals that they do not support Unikernels and that there is room for tailoring the instantiation process to the NFV scenario and to enhance their performance.

This work extends \cite{piervim} and addresses the above question, providing a quantitative analysis of the VIMs under evaluation. Moreover, we release as open source the customized VIMs under analysis complemented with a rich set of tools for their
performance evaluation \cite{ourgitrepo, etsiosm}. The main contributions of this work are the following:
\begin{itemize}
    \item Analysis of the instantiation process of three VIMs: Nomad \cite{nomad}, OpenVIM \cite{etsiosm} and OpenStack with Nova Network and with Neutron Network \cite{openstack};
    \item Description of a general reference model of the VNF instantiation process \optional{,based on the aforementioned analysis};  
    \item Modifications of these VIMs to integrate Xen hypervisor and instantiate ClickOS Unikernels; 
    \item Realization of performance evaluation tools \optional{to measure VM instantiation times, by integrating existing tools (e.g. OpenStack Rally) with our own tools (Nomad Pusher, Packet Analyzers) and by extending VIMs' source code to collect integrative measurements};
    \item Evaluation of ClickOS VMs instantiation times in OpenStack, Nomad and OpenVIM;
    \item Performance tuning of the VIMs to improve instantiation times of ClickOS based VMs;
\end{itemize}
The purpose of the work is not to compare the three VIMs to select the best one, but rather to understand how they behave and find ways to reduce the instantiation times. \optional{We believe that the proposed model for VNF instantiation is applicable to other VIMs not considered here. Starting from the implemented components, experimenters/developers can build solutions for NFV frameworks and extend our tools to experiment on other VIMs.} The rest of the paper is structured as follows: firstly, we give some background information (Section~\ref{sec:background}) of NFV framework and VIMs (Section~\ref{sec:vims}). Section~\ref{sec:modelling} describes the general model of the VNF instantiation process and the specific models for the three considered VIMs. In Section~\ref{sec:modifications}, the modifications needed to boot Unikernels are illustrated. Section~\ref{sec:experimentalresults} provides a performance evaluation of the Unikernels instantiation process. Finally, Section~\ref{sec:relatedwork} reports on related work and in Section~\ref{sec:conclusions} we draw some conclusions and highlight the next steps.

\vspace{-1ex}
\section{Background on NFV architecture}
\label{sec:background}

\begin{figure}
    \centering
    \includegraphics[width=0.485\textwidth]{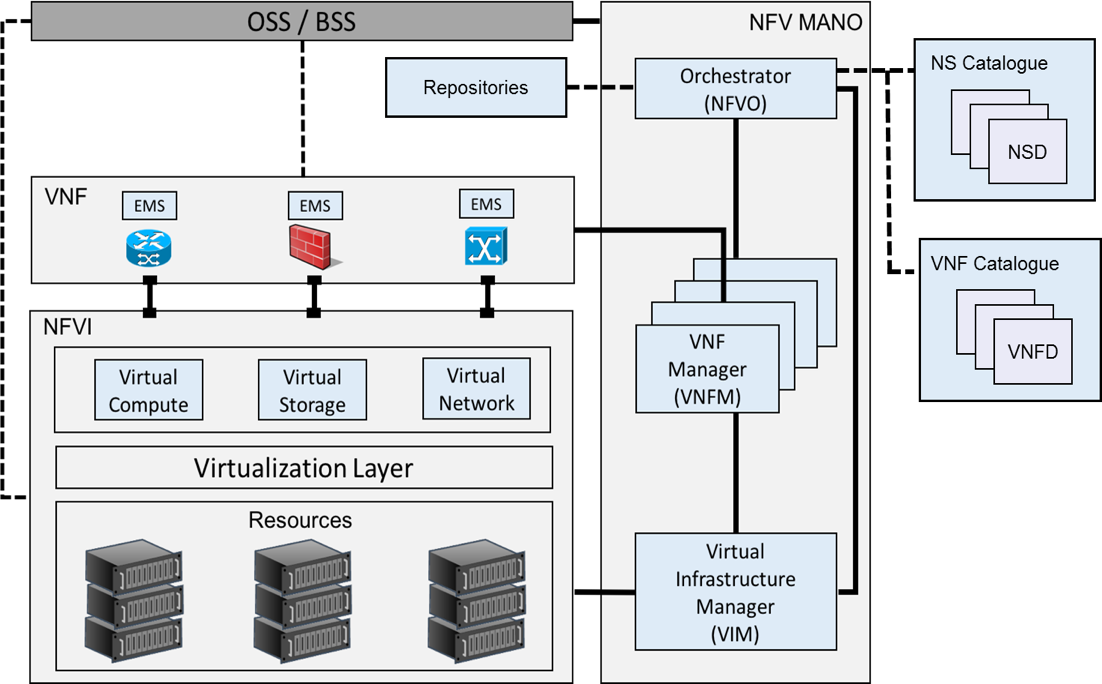}
    \caption{NFV architecture}
    \label{fig:nfv}
    \vspace{-3ex}
\end{figure}

The building blocks of the ETSI NFV~\cite{nfv} and NFV MANO (MANagement and Orchestration)~\cite{mano} architecture have been specified in the last years by the ETSI NFV Industry Specification Group through a number of documents describing the management and orchestration framework for VNFs. Its main components are represented in Figure~\ref{fig:nfv}. The \textit{VNFs} leverage resources provided by the \textit{NFVI} for the execution of the network services. The \textit{NFVI} layer is typically decomposed into compute, networking and storage hardware resources provided by COTS hardware and by the Hypervisor software which implements the \textit{Virtualization layer} abstracting the underlying hardware resources. 

The NFV MANO components represented in the right part of Figure~\ref{fig:nfv} are responsible for the management of the physical infrastructure and the management and orchestration of the \textit{VNFs}. The \textit{NFVI}'s resources are managed by one or more \textit{VIMs}, which expose northbound interfaces to the \textit{VNF Managers} (VNFM) and to the \textit{Network Functions Virtualisation Orchestrator} (NFVO). The VNFM performs the lifecycle management of the \textit{VNFs}. The associated \textit{VNF Catalogue} stores the \textit{VNF Descriptors} (VNFDs), which describe the structural properties of the \textit{VNFs} (e.g. number of ports, internal decomposition in components) and their deployment and operational behaviour. A \textit{VNF} is decomposed into one or more \textit{Virtual Deployment Units} (VDUs) which can be mapped to \textit{VMs} or \textit{Containers} to be deployed and executed over the \textit{NFVI}.
The NFVO is responsible for the overall orchestration and lifecycle management of the \textit{Network Services} (NSs), which combine \textit{VNFs} according to an interconnection topology. A \textit{Network Service} is represented by an \textit{NS Descriptor} (NSD), which captures the relationship between \textit{VNFs}. The NSDs are stored in an \textit{NS Catalogue} and are used by the NFVO during the deployment and operational management of the services. The NFVO is also responsible for the “on-boarding” and validation of the descriptors of \textit{VNFs} and NSs. \optional{Concerning the \textit{VNFs} lifecycle, according to the model defined above, the network operator is able to build NSs combining \textit{VNFs} by means of an orchestrator (NFVO). The NFVO is able to process the descriptors of network services (NSDs) and of \textit{VNFs} (VNFDs) and to interact with the manager (\textit{VIM}) of the \textit{NFVI} in order to deploy the components that implement the service and interconnect them appropriately. In particular, the descriptors of the \textit{VNFs} include the \textit{VDUs}, which are mapped into virtual computing resources that can be deployed over the deployment and execution infrastructure (\textit{NFVI}).} In the aforementioned architecture the VIMs have the key role of interacting with the infrastructure layer in order to manage and orchestrate the virtual resources which are required for the execution of the VNFs.

\vspace{-1ex}
\section{VIMs under evaluation}
\label{sec:vims}

In this work, we consider three general purposes VIMs: OpenStack with Nova network and Neutron network; OpenVIM from the OSM community; and finally Nomad from HashiCorp. Note that we refer to Nomad as a \textit{VIM} while it is also commonly referred to as an \textit{orchestrator}. We do it to be consistent with the ETSI NFV architectural model that separates the Orchestrator (NFVO) from the VIM (Figure~\ref{fig:nfv}).

\textbf{OpenStack} is a Cloud platform designed to manage large-scale computing resources in a single administrative domain, and it aims at providing a solution for both public and private clouds. OpenStack is composed of different sub-projects. Among them, Nova provides the functionality for orchestrating and managing the computing resources. Its architecture envisages a single Nova node and a number of compute nodes. The Nova node schedules the computing tasks and manages the life-cycle of the virtual computing resources, while the compute nodes run locally in the Infrastructure nodes and interact with the local virtualization managers (Hypervisors). Among the other sub-projects in OpenStack, the most important are: Keystone, the identity management service; Glance, the image store; Cinder, the storage service; Nova network and Neutron, the \textit{networking as a service} components; and Horizon, the official dashboard (GUI) of the project. 

As mentioned above two \textit{networking as a service models} exist in OpenStack, providing two different implementations of the same service. The former, Nova network, is the legacy model and is implemented as a sub-process embedded in the Nova project. Even if Nova network is now deprecated, it is still used in several deployments. Neutron, the new networking as a service model, moves the network functionality out of Nova, making the projects self-contained, less-dependent on the latter and simplifying the development process. It resolves some architectural flaws by introducing a technology agnostic layer and enabling extensions through a plugins oriented framework. Moreover it provides more functionalities to the end-users: complex networking topologies that go beyond the flat models and the VLAN-aware model, better multi-tenancy with the support of the tunneling protocols, load balancing, firewall services and many others.

\textbf{Nomad} by HashiCorp is a minimalistic cluster manager and job scheduler, which has been designed for micro services and batch processing workloads. Once compiled, it is a self-contained executable which provides in a small binary all the functionality of a resource manager and of a scheduler. \optional{Nomad can work both in multi-datacenter and multi-region scenarios.} In the Nomad architecture, there are two types of nodes: Servers and Clients. The Servers take care of scheduling decisions. \optional{all the Server nodes participate in scheduling decisions.} The Clients are the resource managers and locally run the Jobs submitted by the Servers (the Jobs are the unit of work in Nomad). \optional{they are composed of one or more Task Groups, which are themselves collections of Tasks}.

\begin{figure}
    \centering
    \includegraphics[width=0.485\textwidth]{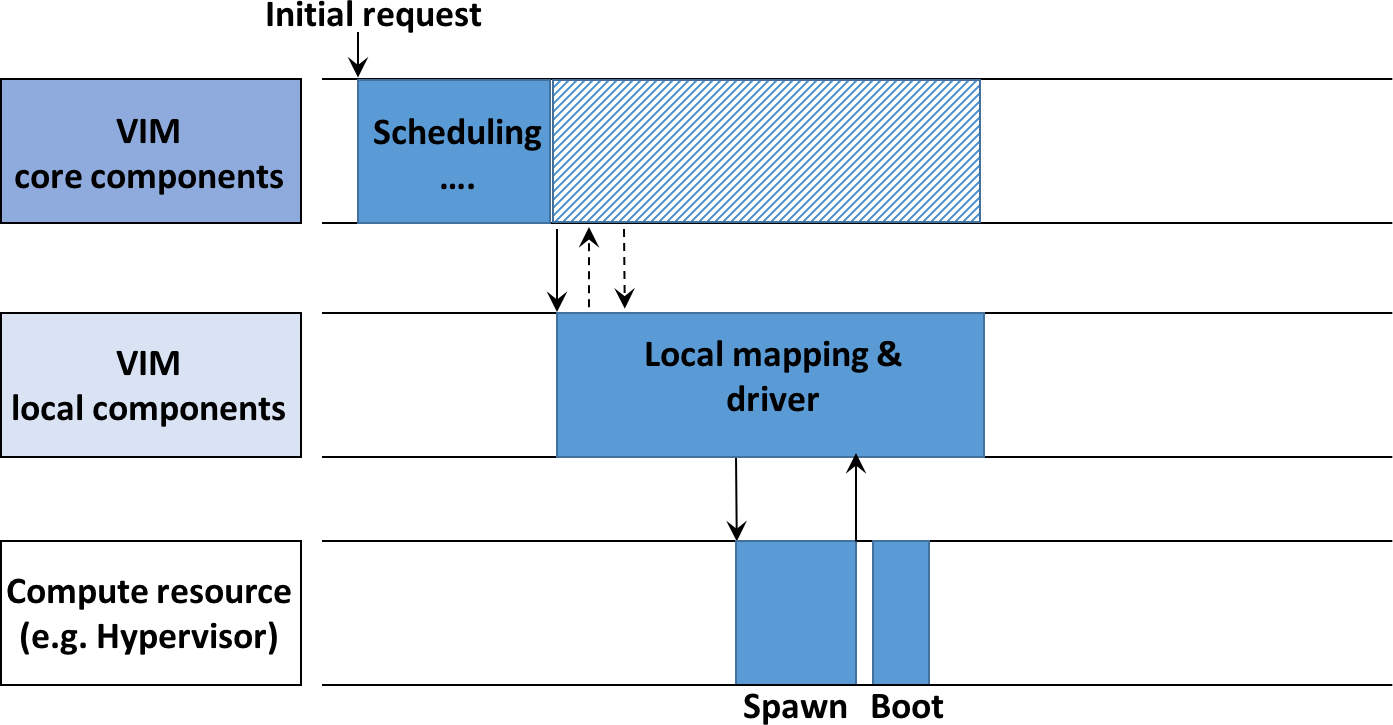}
    \caption{VIM instantiation general model}
    \label{fig:vimGeneral}
    \vspace{-3ex}
\end{figure}

\textbf{OpenVIM} is the reference implementation of an NFV VIM in the context of the ETSI MANO architecture. Originally developed as VIM of the OpenMANO suite, it is now maintained under the umbrella of the OSM project managed by ETSI. OpenVIM is designed as a single software component which is able to manage the entire life-cycle of the virtual infrastructure. Its internal data-model follows NFV recommendations~\cite{nfvper}. On the northbound side it offers an OpenStack-like REST API interface enhanced to be compatible with the NFV MANO framework, while on the southbound side it interacts with the infrastructure (compute nodes) providing the physical resources for VNF instantiation. The networking component assumes a flat network topology, with the possibility of segregating tenants by means of VLANs. \optional{VMs can be interconnected in three modes: Bridging, SDN controller and Open vSwitch controller. In the first mode OpenVIM leverages \textit{libvirt} to interconnect co-located VMs by Linux Bridge or by OVS working on legacy switch mode. The second mode leverages an SDN controller and \textit{OpenFlow}-enabled network to establish communication between VMs. Finally, the third mode introduces VXLAN tunnels between OVS switches. The networking is not totally delegated to the SDN control plane since OpenVIM connects VMs on same compute node through bridges. Then, the SDN controller manages \textit{OpenFlow} switches which interconnect the compute nodes. } The interaction with the NFVI is done by means of \textit{SSH} connections which transport the instructions for \textit{libvirt}, which in turn controls the hypervisors. At the time of writing only the KVM/Qemu hypervisor is natively supported by the OpenVIM project.

There are important differences between the VIMs under evaluation; OpenStack is a complete and mature Cloud suite composed of 11 core projects (without considering deployment tools) and a number of side components for a total of 46 projects. It provides a large set of functions that are essential for managing a Cloud infrastructure. Networking components are very mature, they support a wide range of technologies and they have built-in integration with several SDN controllers which are suitable for the NFV use case. At the same time, this wide range of functionalities and sub-projects can add unnecessary overhead when employed in the NFV context.
A different architectural approach is taken by the Nomad VIM: it is packed in a single executable with a small footprint (about 30 MB), and it is focused on a minimal complete functional set. Thus, it only contains what is strictly necessary to schedule the virtual computing resources and to instantiate them in an infrastructure node. In-between the two approaches described so far, OpenVIM provides a minimalistic VIM implementation targeting the NFV use case and represents the reference implementation for ETSI NFV. However, it is still in its infancy and it does not reach the same level of maturity of OpenStack. Moreover, its future is still not clear inside the OSM community.

OpenStack and OpenVIM share an advanced networking component and they do have support for an SDN ecosystem. Concerning the development of new functionality, it is easier to provide extensions to Nomad and OpenVIM, being smaller projects when compared to OpenStack. A common characteristic between the VIMs under analysis, is that they do not focus on a specific virtualization technology, but provide an extensible framework to support different types of virtual computing workloads. Instead, other projects, such as Kubernetes~\cite{kubernetes}, are designed for a specific virtualization technology (in this case \textit{containers}), and modifying them to support the instantiation of Unikernels or VMs would require changes to the software design and code base.
\vspace{-1ex}
\section{Modelling Approach}
\label{sec:modelling}

The proposed general model of the VM instantiation process is shown in Figure~\ref{fig:vimGeneral}. We decompose the operations among the \textit{VIM core components}, the \textit{VIM local components} and the \textit{Compute resource/hypervisor}. The VIM core components are responsible for receiving the VNF instantiation requests (i.e., the \textit{initial request} in Figure~\ref{fig:vimGeneral}) and for choosing the resources to use, i.e., the \textit{scheduling}. This decision is translated into a set of requests which are sent to the \textit{VIM local components}. These are located near the resources and are responsible for enforcing the decisions of the \textit{VIM core components} by mapping the received requests to the corresponding technology API calls. These APIs are typically wrapped by a \textit{driver} or, borrowing from the OpenStack terminology, by a \textit{plugin}, which is responsible for instructing the Compute resource to instantiate and boot the requested VNFs and their resources.

\begin{figure}[t]
    \centering
    \includegraphics[width=0.485\textwidth]{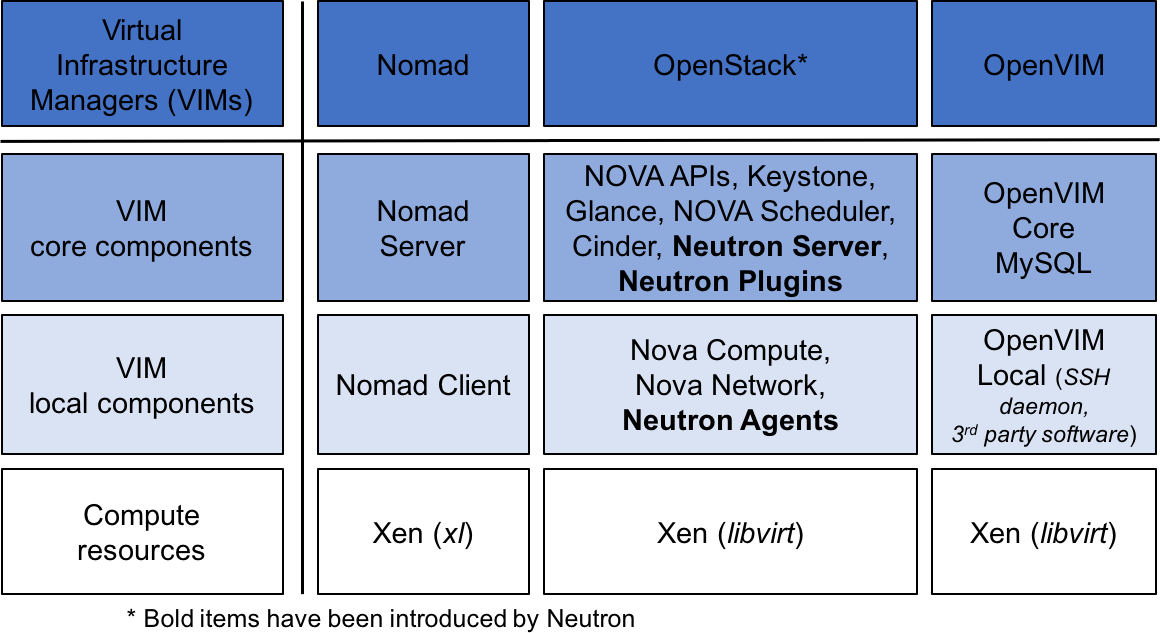}
    \caption{Mapping of the reference model to the VIMs}    
    \label{fig:vimMapping}
    \vspace{-3ex}
\end{figure}

Figure~\ref{fig:vimMapping} shows how the OpenStack, OpenVIM and Nomad components can be mapped to the proposed reference model. In the next subsections, we provide a detailed analysis of the operations of the considered VIMs. Two mappings have been presented for OpenStack due to the big architectural differences introduced by Neutron. The new standard OpenStack deployment uses Neutron as networking component, while the so-called OpenStack Legacy version uses Nova. As for the Compute resource/hypervisor, in this work we focus on Xen~\cite{xen}, an open-source hypervisor commonly used as resource manager in production clouds. Xen can use different \textit{toolstacks} (tools to manage guests creation, destruction and configuration); while Nomad uses \textit{xl}, the others use \textit{libvirt}. Regarding networking, we did not consider the offloading to an SDN controller and we used the built-in plugins of the VIMs.

\subsection{Modelling OpenStack Legacy}
\label{sec:openstacklegacymodeling}

Figure~\ref{fig:vimNova} shows the scheduling and instantiation process for OpenStack Legacy. The requests are submitted to the Nova API using the HTTP protocol (REST API). The Nova API component manages the \textit{initial requests} and stores them in the Queue Server. At this point, an authentication phase towards Keystone is required. The next step is the retrieval of the image from Glance, which is required for the creation of virtual resources. At the completion of this step, the Nova Scheduler is involved: this component performs tasks scheduling by taking the requests from the Queue Server, deciding the Compute nodes where the guests should be deployed and sending back its decision to the Nova API (passing through the Queue Server). The components described so far are mapped to the \textit{VIM core components}, in our model. After receiving the scheduling decision from the Nova Scheduler, the Nova API contacts the Nova Compute node. This component manages the interaction with the specific hypervisors using the proper \textit{toolstack} and can be mapped (along with Nova Network) to the \textit{VIM local components}. The VM instantiation phase can be divided in two sub-steps: Network creation and Spawning. Once this task is finished, Nova Compute sends all the necessary information to libvirt, which manages the spawning process instructing the Xen hypervisor to boot the virtual machine. When the completion of the boot process is confirmed, Nova Compute sends a notification and the Nova API confirms the availability of the new VM. At this point the machine is ready and started. The above description, reflected in Figure~\ref{fig:vimNova}, is a simplified view of the actual process: for sake of clarity many details have been omitted. For example, the messages exchanged between the components traverse the messaging system (Nova Queue Server, which is not shown), and at each step the system state is serialized in the Nova DB (not shown).

\subsection{Modelling Openstack (with Neutron)}
\label{sec:openstackneutronmodelling}

\begin{figure*}[!t] 
    \centering
    \subfloat[VIM instantiation model for OpenStack Legacy]{%
    \includegraphics[width=0.485\textwidth]{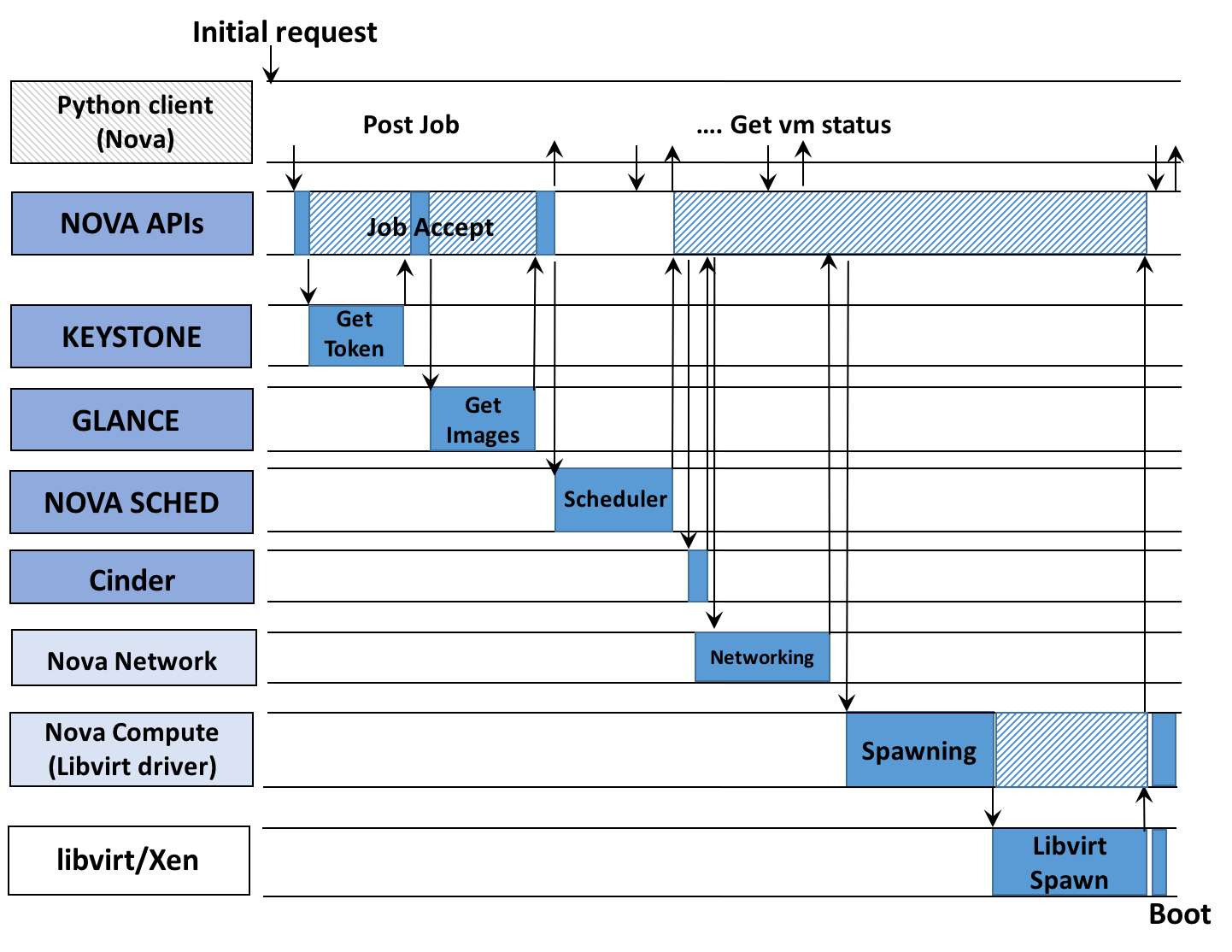}
    \label{fig:vimNova}}\hfill
    \subfloat[VIM instantiation model for OpenStack (with Neutron)]{%
    \includegraphics[width=0.485\textwidth]{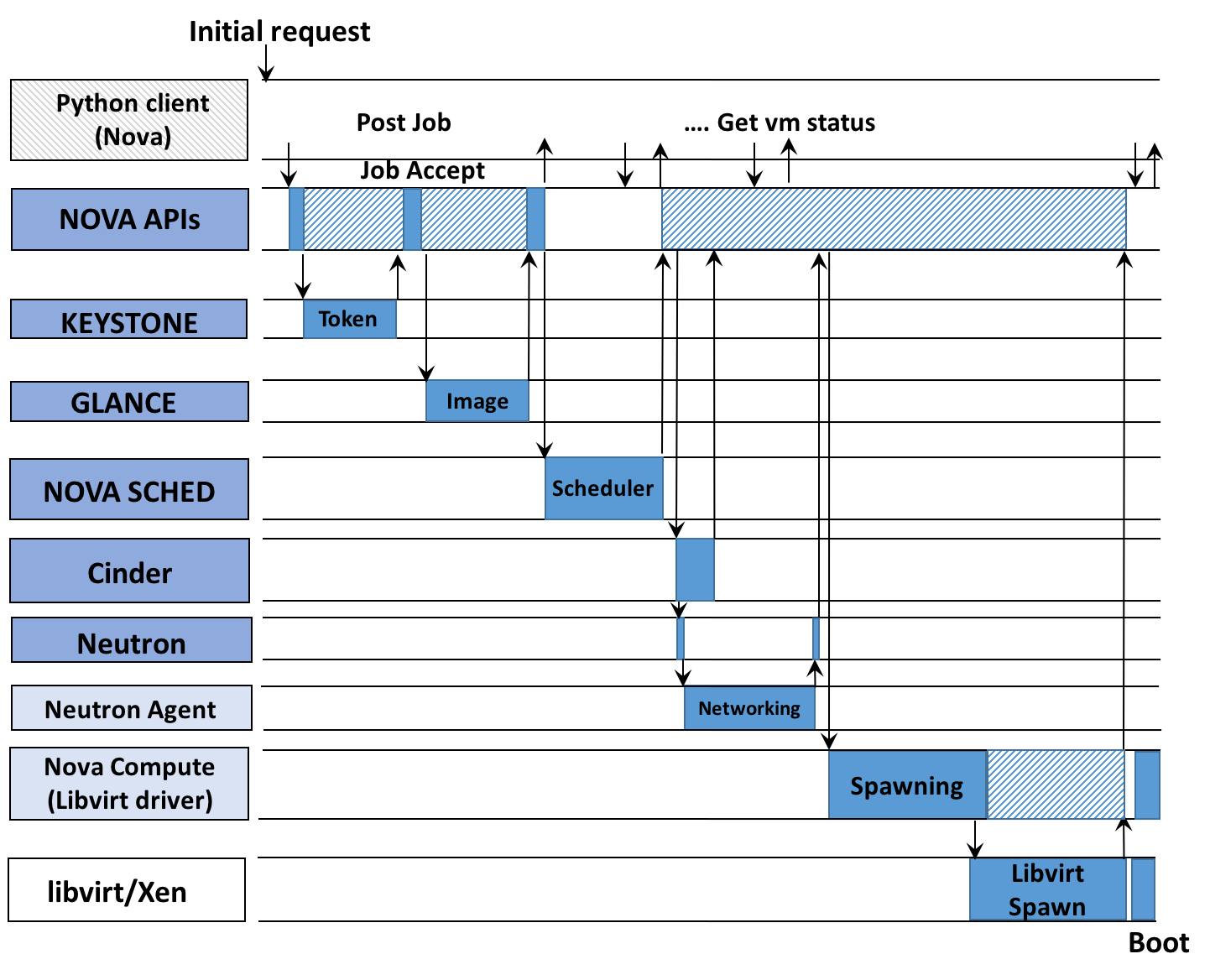}
    \label{fig:vimNovaNeutron}}
    \vspace{-1ex}
    \caption{VIM instantiation models for OpenStack}
    \vspace{-3ex}
\end{figure*}

The VM instantiation process in OpenStack is similar to the one described above for OpenStack Legacy, but relies on Neutron for the network management. Provisioning a new VM instance involves the interaction between multiple components inside OpenStack. The main components involved in the instantiation process are: Keystone, for authentication and authorization; Nova (with its subcomponents Nova API, Nova Scheduler and Nova Compute), for VM provisioning; Glance, the image store component; Cinder, to provide persistent storage volumes for VM instances; Neutron, which provides virtual networking, and which is split in the two subcomponents Neutron controller (\textit{VIM core component}) and Neutron agent (\textit{VIM local component}). Figure~\ref{fig:vimNovaNeutron} shows the main phases of the scheduling and instantiation process for OpenStack Nova, the main differences are summarized hereafter. 

Once Nova API receives back the image URI and its metadata from Glance, and the host id from Nova Scheduler; next step is the interaction with Cinder for the retrieval of the block storage. First difference we can appreciate is a more advanced scheduling function: it selects the hosts via filtering and weighting instead of using simple round robin schemes. In parallel, Nova API requests Neutron (\textit{VIM core component}) to configure the network for the instance. At this point the workflow is managed by the OpenStack \textit{VIM local components}: Neutron server validates the request and then instructs the local Agent to create and properly set-up the network for the VM. As soon as the Agent finishes the creation Nova APIs gets the relevant information and contacts Nova computes for the spawn of the VM. Finally, the workflow follows the same step as of OpenStack Legacy.

\subsection{Modelling Nomad}

\begin{figure}
    \centering
    \includegraphics[width=0.485\textwidth]{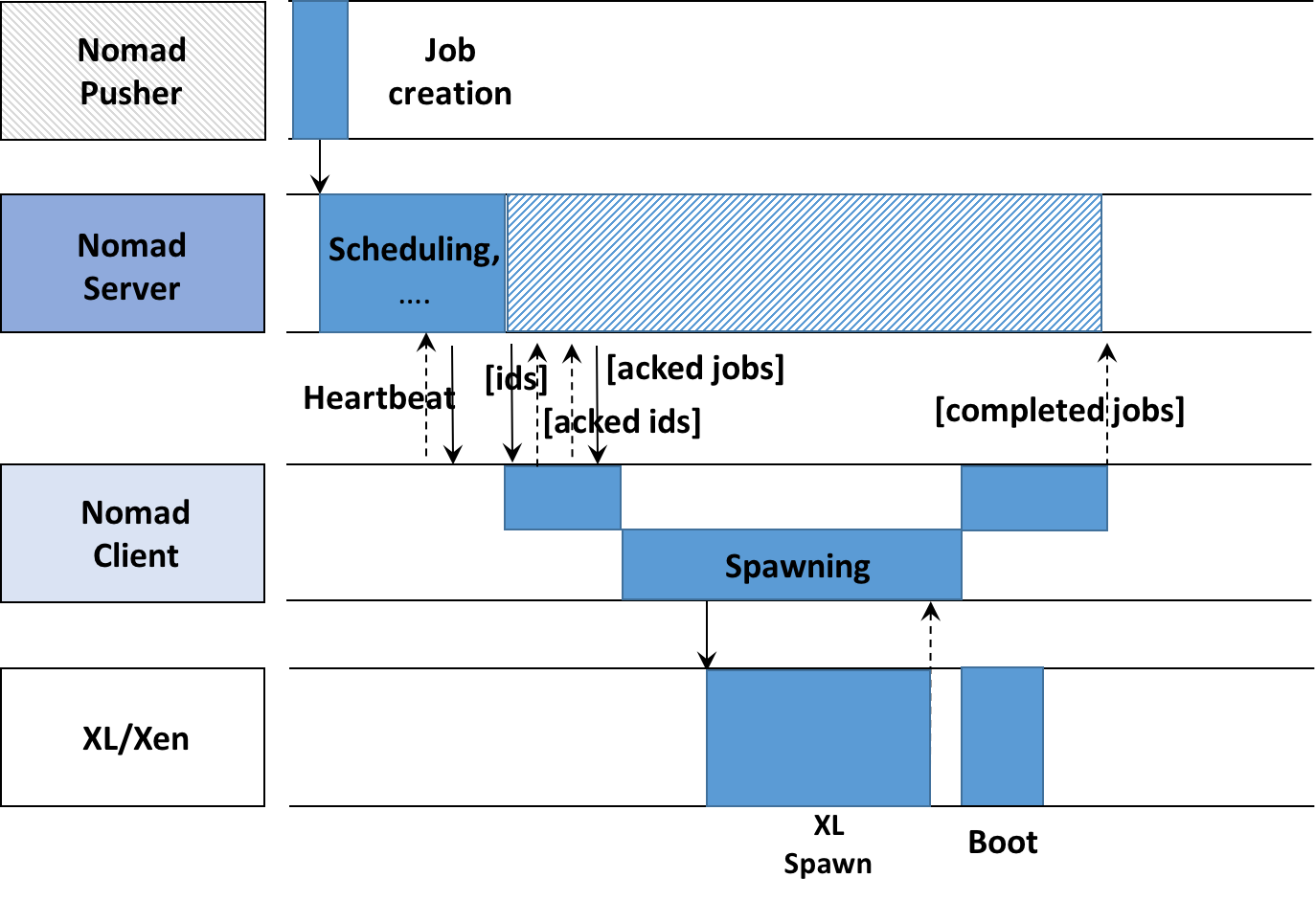}
    \caption{VIM instantiation model for Nomad}
    \label{fig:vimNomad}
    \vspace{-3ex}
\end{figure}

The scheduling and instantiation process for Nomad is shown in Figure~\ref{fig:vimNomad}. According to our model (cf.\ Figure~\ref{fig:vimMapping}), the Nomad Server is mapped to the \textit{VIM core components}. It receives the requests for the instantiation of VMs (jobs) through the REST API. Once the job has been accepted and validated, the Server takes the scheduling decision and selects which Nomad Client node to run the VM on. The Server contacts the Client sending an array of job IDs.  As response the Client provides a subset of IDs which are the ones that will be executed in this transaction. The Server acknowledges the IDs and the Client executes these jobs. The Nomad Client is mapped to the \textit{VIM local components} and interacts with Compute resources/hypervisors. In Xen, this interaction is done via a user-level \textit{toolstack}, of which Nomad supports several. We used xl (see section \ref{sec:modifications}), the default Xen toolstack, to interface with the local Xen hypervisor. xl provides a command line interface for guest creation and management. The Client executes these jobs loading the Nomad Xen driver, which takes care of the job execution interacting with the xl \textit{toolstack}. The instantiation process takes place and, once completed, the Client notifies the Server about its conclusion. Meanwhile the boot process of the VM starts and continues asynchronously with respect to the Nomad Client. 

\subsection{Modelling OpenVIM}

Borrowing the terminology we have defined in Figure~\ref{fig:vimMapping}, OpenVIM Core and OpenVIM DB compose the \textit{VIM core components}. The submission of tasks (i.e., commands and associated descriptors) can be performed through the OpenVIM Client. The requests from the client are forwarded to the core via REST API. Each task request is then mapped to a specific image flavor and image meta-data. These, together with the status of the Compute nodes (retrieved from the OpenVIM DB) are given as input to the scheduling process, which decides the Compute nodes that will handle these requests. The taken decision is written to the database and the OpenVIM Client is notified returning a HTTP 200 OK message. At this point the task is split into a network request and an instantiation request. If the networking part is managed through a compatible SDN controller, the network request is forwarded to a network thread which takes care of interacting with SDN controller of the infrastructure in order to perform all the networking related tasks. Otherwise, the network request is directly forwarded to libvirt. Instead, the instantiation is performed partially on the controller node and partially on the local node: OpenVIM Core is a multi-thread application where each Compute node is associated with a thread and a work-queue. According to the scheduling decision, the jobs are submitted to the proper work queue and subsequently the spawning process can start. The thread starts this phase by generating the XML description of the instance and submitting it to the OpenVIM Local's libvirt daemon running on the chosen compute node. The requests, data and what is needed for the spawning is forwarded to the destination leveraging SSH connections. If needed, 3rd party software can be invoked for further configurations (e.g., Open vSwitch). At this point the job is completely offloaded to the Compute resources component. The libvirt daemon creates the instance based on the received XML. After this step, the associated thread on the OpenVIM Core calls the libvirt daemon again to instruct it to start the previously-created instance. Finally, the libvirt daemon boots the instance through the Xen hypervisor and the information on the started instance is updated in the OpenVIM Database. The OpenVIM Client can obtain up to date information during the VM lifetime by polling its status. Figure~\ref{fig:vimOpenVIM} shows the instantiation process in OpenVIM. For sake of clarity, networking details have been omitted.

\vspace{-1ex}
\section{VIM Modifications to boot Micro-VNFs}
\label{sec:modifications}

We designed and implemented some modifications in the considered VIMs to enable the instantiation of Micro-VNFs based on ClickOS using OpenStack, Nomad and OpenVIM. VIMs offers support for various type of virtualization: paravirtualization (PV), Hardware Virtualization (HVM) and container-based virtualization. Unikernels can both run fully virtualized or paravirtualized using different hypervisors (e.g., KVM\cite{kvm}, XEN\cite{xen}). However, this does not imply that all VIMs support Unikernel based VNFs. The main reasons for these adaptations lie in the peculiarities of the ClickOS VMs compared to regular VMs. 

A regular VM can boot its OS from an image or a disk snapshot that can be read from an associated \textit{block device} (disk). The host hypervisor instructs the VM to run the boot loader that reads the kernel image from the block device. A small variation of this process boots the VM with a special bootloader (provided by the hypervisor). The latter starts the boot cycle and then loads the kernel specified in the VM bootloader config file (e.g., \textit{menu.lst}) on the VM image. On the other hand, we are interested in ClickOS based Micro-VNFs, provided as a tiny self-contained kernel, without a block device (boot executing directly the kernel). These VMs need to boot from a so-called \textit{diskless image}. Note that Unikernel VMs, depending on the function that they perform, can also use a block device for storage purpose but this device is not needed to boot the VM. When instantiating such VMs, the host hypervisor reads the kernel image from a file or a repository and directly injects it in the VM memory, meaning that the modifications inside the VIM can be relatively small and affecting only the component interacting with the virtualization platform to boot \textit{diskless images}. We define as \textit{stock} the VIM versions that only include the modifications required to boot the ClickOS VMs. We analyzed the time spent by the \textit{stock} VIMs during the different phases of the driver operations. Then, taking into account the results, we proceeded with an optimization trying to remove those functions/operations not directly needed to instantiate Micro-VNFs. We define as \textit{tuned} these optimized versions of the orchestrators. 

From the point of view of the development effort, it has been possible to realize the VIMs modifications and maintain them updated through the further releases with an overall effort in the order of 3.5 Person Months (PM). We roughly spent 1.5 PM for OpenStack, 1.5 PM for OpenVIM (including the effort to get the changes merged in the upstream repository of the OSM project \cite{etsiosm}) and 0.5 PM for Nomad. The amount of work to implement the modifications for each VIM is heavily dependent on the specific project: for example for Nomad and OpenVIM it was relatively easy to introduce the support for booting from \textit{diskless images} thanks to their simple architectures and to the flexibility of their \textit{southbound} framework. In contrast, it was more difficult to understand how to modify OpenStack, due to its larger codebase, supporting several types of hypervisors, each through more than one toolstack. In our previous work~\cite{piervim}, we implemented a ``hack'' to instantiate ClickOS images, while in this work we have devised an alternative solution, cleaner and easier to integrate in the main OpenStack development line. Our patches to OpenStack, Nomad and OpenVIM are available at \cite{ourgitrepo} and in the OSM repository. The following subsections elaborate on the realized solutions and provide some implementation details.

\subsection{OpenStack Legacy -- The ``hacky'' solution}
\label{sec:legacymod}

In our previous work~\cite{piervim}, for OpenStack Legacy we enabled the boot of \textit{diskless images} targeting only one component (Nova Compute) and a specific toolstack (\textit{libvirt}). We selected this toolstack because it supports \textit{libxl}, which is the default Xen toolstack API. We first implemented the minimal set of changes to the \textit{libvirt} driver as needed to instantiate the ClickOS VMs. In particular, we modified the XML description of the guest domain provided by the driver: by default \textit{libvirt} can support different domain types, i.e., virtualization technologies. ClickOS requires the paravirtualization technology provided by the Xen hypervisor. The proposed approach involved changing the XML description on the fly, before the actual creation of the domain. The change, required to enable the boot through \textit{libvirt}, consists in using paravirtualization and providing the ClickOS image as the kernel. After this change we need a way to trigger our code. We did not patch the OpenStack image store (\textit{Glance}) because a mainstream patch was out of the scope of our previous work. We resorted to a simple hack to trigger the execution of our code, adding a new image to the store with a specific image name. When this specific name was selected for instantiation, our patches to the Nova compute \textit{libvirt} driver were executed and the ClickOS image was loaded directly from the disk (not from Glance). Moreover we limited the interaction with Cinder since the targeted Unikernel does not need a disk to boot up. Note that the above modification significantly changed the workflow: in OpenStack it is possible to instantiate VMs with a 0-byte disk, but even in this case, a volume is mandatory. In fact, when the VM is created, it needs to be associated with an ephemeral root disk that will contain the bootloader and core operating system files. In the deployment of OpenStack with Neutron we have introduced back this step as explained in following Section (\ref{sec:openstackmod}).

\begin{figure}
    \centering
    \includegraphics[width=0.485\textwidth]{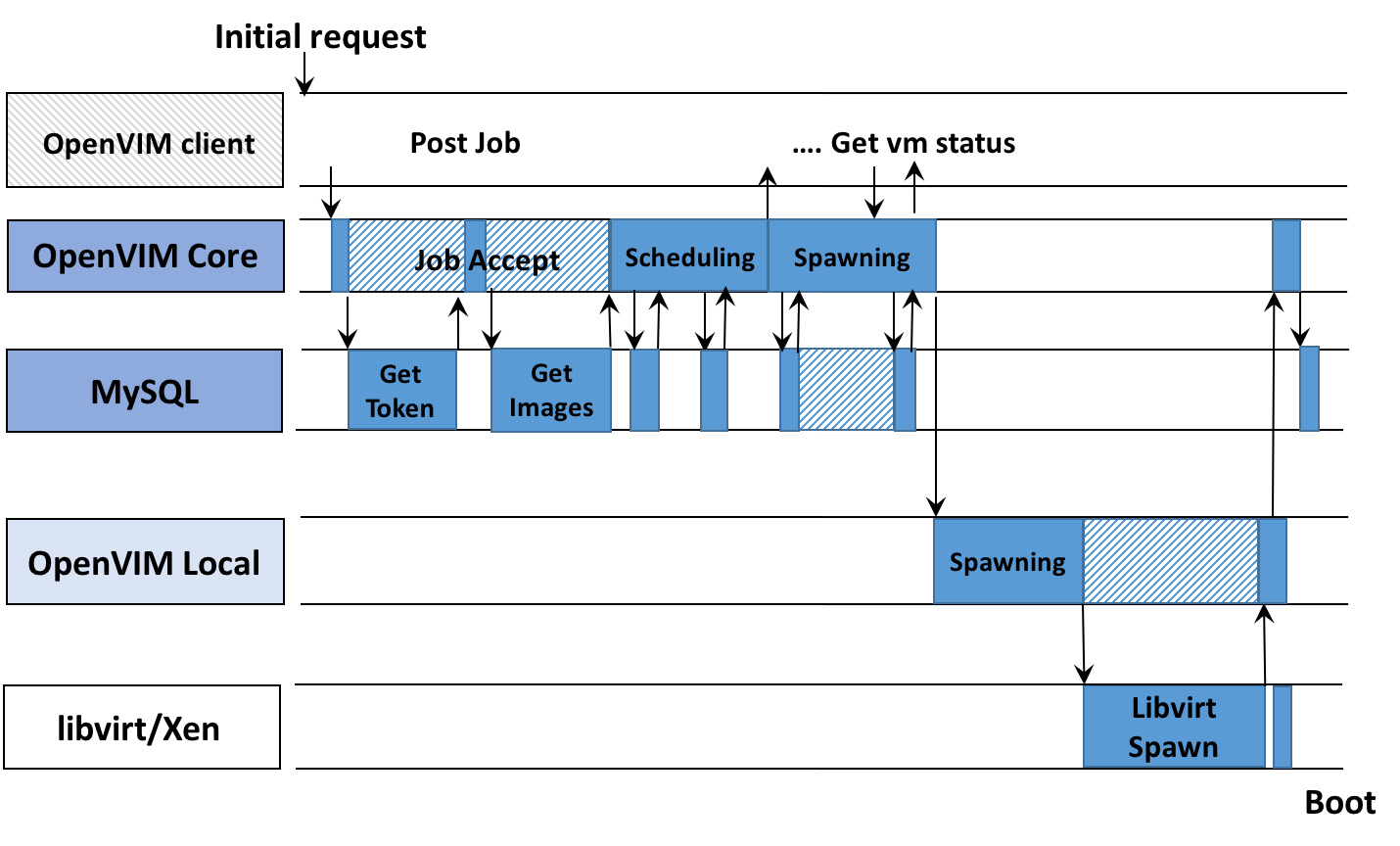}
    \caption{VIM instantiation model for OpenVIM}
    \label{fig:vimOpenVIM}
    \vspace{-3ex}
\end{figure}

\subsection{OpenStack -- Borrowing the AKI approach}
\label{sec:openstackmod}

Concerning OpenStack with Neutron networking, we have implemented an improved modification starting from the work described in the previous subsection. We have reused part of the modifications done to the \textit{libvirt} driver in the \textit{tuned} version of OpenStack. Instead for the \textit{stock} version, we have improved the patch introducing directly the support for ClickOS images. We have extended the OpenStack image store (\textit{Glance}) leveraging the OpenStack support for Amazon (AWS) EC2 images. Indeed, Amazon EC2 supports paravirtualization, and an Amazon image is typically composed by three different objects: i) \textit{AMI}, the Xen DomU guest operating system; ii) \textit{ARI}, the Xen paravirtualized ramdisk; iii) \textit{AKI}, the Xen paravirtualized kernel. Going into details, an AKI is a pre-configured bootable mini kernel image which is pre-built and provided by Amazon to boot instances. It contains a specialized bootloader for the \textit{Xen} environment required to boot an instance from an AMI. In particular, the \textit{AKIs} are shipped with the \textit{PV-GRUB} bootloader which is derived from MiniOS~\cite{stubdomains}. Then, we created a new kernel image called \textit{UKI} (Unikernel Image) and we used ClickOS in the image instead of \textit{PV-GRUB}. We added new Amazon EC2 images in \textit{Glance} containing ClickOS as AKI and then we excluded the AMI and ARI. In this way, we are able to load the images directly from the store without employing any hack. Moreover, the patch is likely to be accepted by OpenStack community since it results more compliant to the typical workflow.

\subsection{Nomad Modifications}
\label{sec:nomadmod}

In our previous work, we leveraged the Nomad's support for different types of workload through the \textit{driver} framework. In particular the following virtual resource types have built-in support: \textit{Docker}, \textit{Java VMs}, \textit{QEMU/KVM} and \textit{RKT} engine. Starting from the \textit{QEMU/KVM} support, we developed a new Nomad \textit{driver} for Xen, called \textit{XenDriver}. The new \textit{driver} communicates with the \textit{XL} toolstack and it is also able to instantiate a ClickOS VM. By extending the \textit{driver} framework and changing the job type in the description we added support in Nomad for Xen type jobs without the need to modify the Server: this component is not aware of the details and verifies only the availability of resources against the candidate Clients (Compute hosts supporting the Xen type jobs). The drivers are loaded at boot time and the supported drivers are communicated to the Servers in the context of the Client's registration. Therefore, using the standard Nomad CLI or a HTTP client we can create Xen jobs and submit them to the Server.

\subsection{OpenVIM Modifications}
\label{sec:openvimmod}

For what concerns OpenVIM, we were able to add support for Unikernels with a relatively small change: OpenVIM by default uses \textit{libvirt} as toolstack with KVM/QEMU as hypervisor. First, we have extended the \textit{libvirt} back-end by adding support for the proper libvirt (XML) descriptors. The modification is very similar to the one described in Section~\ref{sec:legacymod}. Moreover, we changed the descriptors to use Xen as hypervisor.

Then the second step of the modification consisted in making aware the \textit{OpenVIM core} of this new type of VMs. This modification has been designed to guarantee backward compatibility with the original OpenVIM supported VM types. We have extended the descriptor formats of the VMs and of the Compute Nodes. For the former, we have introduced a \textit{hypervisor} attribute indicating the technology: \textit{kvm}, \textit{xen-unik} and \textit{xenhvm} are possible values and reflect respectively the original mode, Unikernel mode and full VM mode for Xen. \textit{OSImageType} is the second attribute added for the VMs indicating the type of image to be used. Its presence is mandatory when \textit{xen-unik} is selected as hypervisor\footnote{Currently only the \textit{clickos} value is supported but the \textit{OSImageType} attribute allows for the future support of different types of Unikernels.}.
Regarding the Compute node descriptors, we have added a new attribute describing the type of supported hypervisors. Finally, we have changed the scheduling to take into account the newly introduced attributes: the Compute nodes are now selected based not only on the available resources but also considering the type of supported hypervisors. With these modifications in place we are able to generate the proper \textit{libvirt} (XML) configuration for ClickOS and instantiate and start Unikernels in OpenVIM.

All the changes mentioned above have been recently accepted by the project maintainers and are available from the release \textit{THREE} of OpenVIM.

\subsection{Applicability of the modifications}
\label{sec:applicability}

In this work we have focused on just one Unikernel, namely ClickOS. All the extensions we have realized are applicable to all the Unikernels derived from MiniOS, such as ClickOS and the recent Unikraft/Unicore family. As regards other Unikernels, the proposed modifications would be easily applicable if: i) the Unikernel support Xen Para-Virtualization; ii) the Unikernel can be booted executing directly the kernel; iii) the Unikernel can run with an ephemeral root block device. For example let us consider two other Unikernels: OSv \cite{osv} and Rumprun \cite{rumprun}. The former could be booted leveraging the stock version of OpenStack and OpenVIM and using the extended version of Nomad; but fast-instantiation and tuning cannot be achieved because it does not support para-virtualization, it cannot be instantiated executing directly the kernel and it needs a block device. On the contrary, for the Rumprun Unikernels we can apply in principle all the modifications we have realized since it does not require a disk and supports only the Xen para-virtualization. 
\vspace{-1ex}
\section{Experimental Results and Tuning}
\label{sec:experimentalresults}

In order to evaluate the VIM performances during the VM scheduling and instantiation phases, we combine different sources of information. As a first step, we analyze the message exchanges between the VIM components in order to obtain coarse information about the beginning and the end of the different phases of the VM instantiation process. The analysis of messages is a convenient approach as it does not require modification of the source code. To this aim, we have developed a \textit{VIM Message Analyzer} tool in Python using the Scapy~\cite{scapy} library. The analyzer (available at \cite{ourgitrepo}) is capable of analyzing OpenStack and Nomad message exchanges. Regarding OpenVIM, a separate discussion is needed: we do not use the aforementioned approach because of the high instability of the performance (detailed later in Section~\ref{sec:openvimresults}) we faced during preliminary tests and because in the software architecture there is no strong partition in sub-components running on separate machines: basically OpenVIM is a monolithic multi-thread VIM which interacts directly with \textit{libvirt} running on the compute node. Thus, it has been necessary an accurate breakdown of the different phases and of their timings. For this we have relied on timestamps also on the logs provided by Xen. More details are provided in Subsection~\ref{sec:openvimresults}. For a detailed breakdown of the spawning phase, we have inserted timestamp logging instructions into the code of the Nomad Client and Nova Compute nodes. We generate the workload for OpenStack using Rally~\cite{osrally}, a well known benchmarking tool. For the generation of the the Nomad workload, instead, we have developed the \textit{Nomad Pusher} tool. It is an utility written in the GO language, which can be employed to programmatically submit jobs to the Nomad Server. For what concerns OpenVIM, we leverage the OpenVIM Client CLI for the submission of the jobs to the system. We have executed several experiments to evaluate the performance of the considered VIMs. The complete data set of our experiments is available at \cite{vimdataset}. 

\begin{figure*}[!t] 
  \centering
  \subfloat[Instantiation time breakdown on OpenStack]{%
  \includegraphics[width=0.485\textwidth]{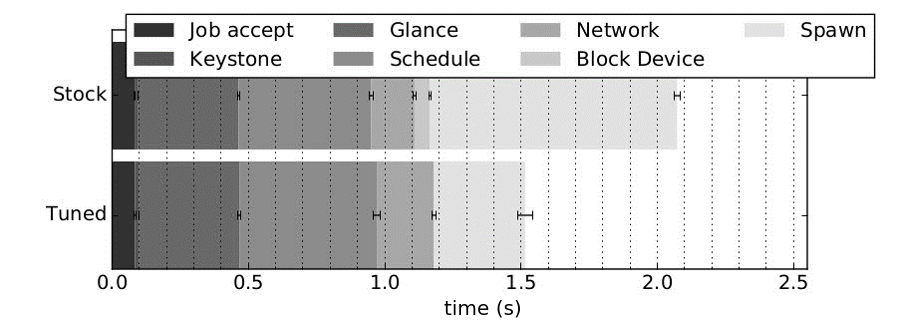}
  \label{fig:osneutronclickos}}\hfill
  \subfloat[Spawning time breakdown on Nova-compute]{%
  \includegraphics[width=0.485\textwidth]{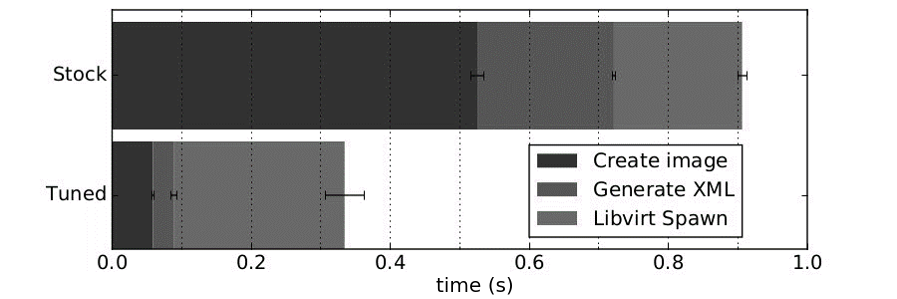}
  \label{fig:neutronclickos}}\\
  \vspace{-2ex}
  \subfloat[Instantiation time breakdown on Nomad]{%
  \includegraphics[width=0.485\textwidth]{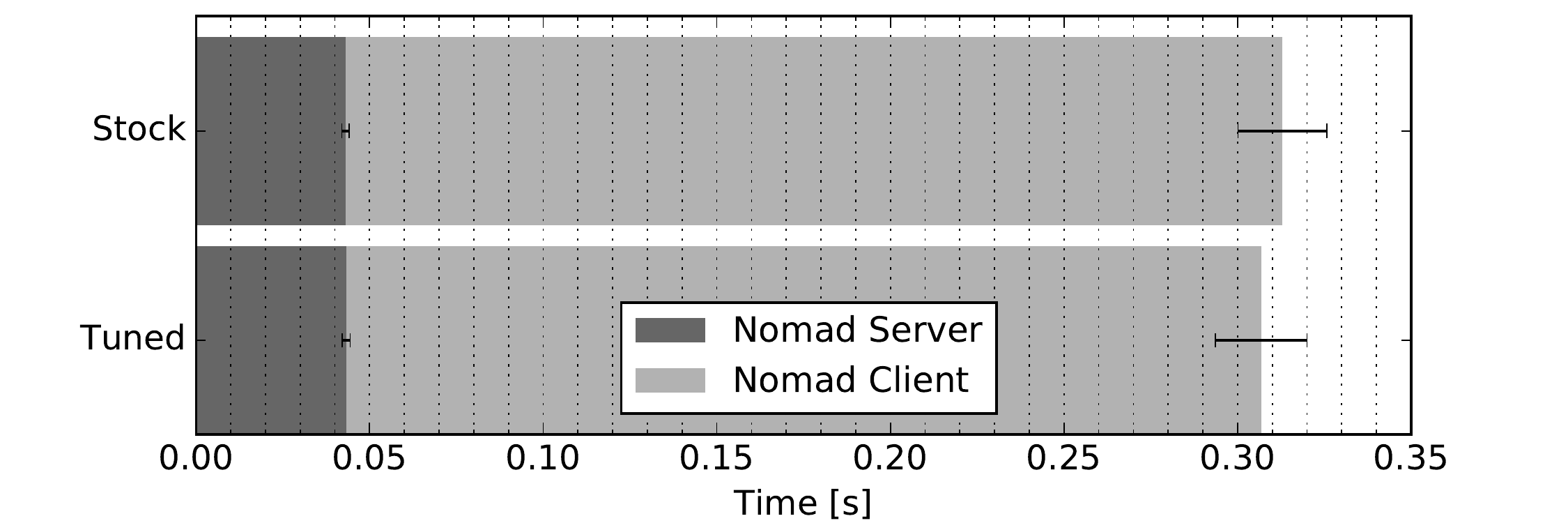}
  \label{fig:nomad_1_clickos}}\hfill
  \subfloat[Spawning time breakdown on Nomad]{%
  \includegraphics[width=0.485\textwidth]{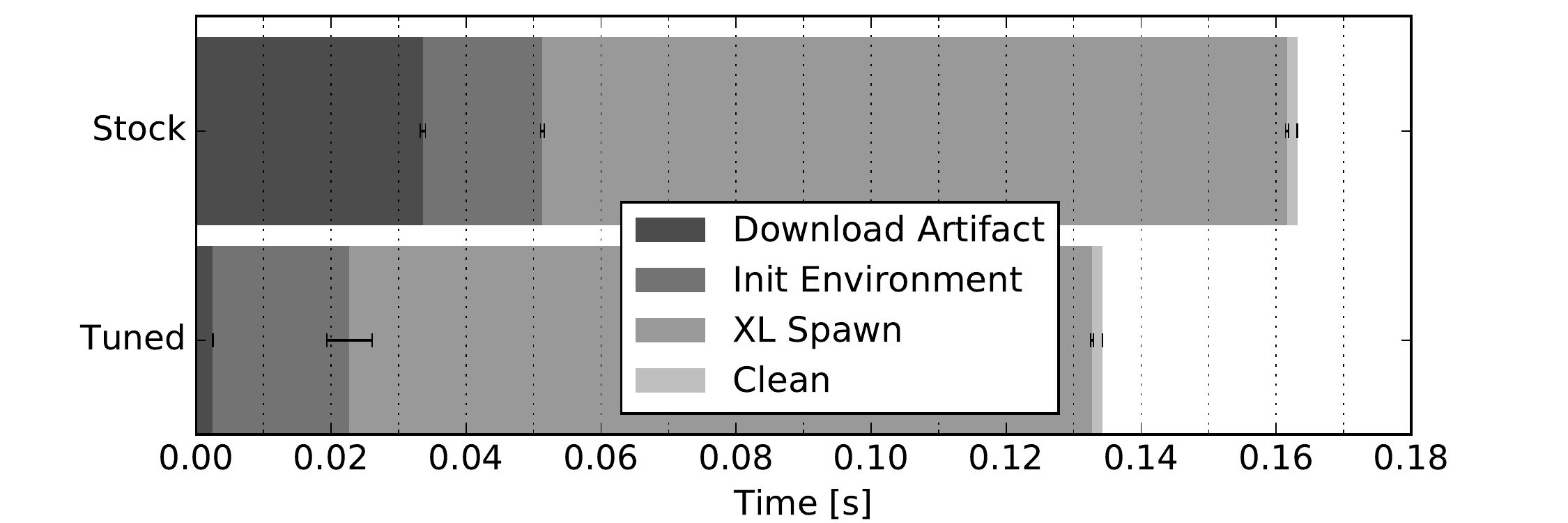}
  \label{fig:nomad_2_clickos}}
  \caption{ClickOS instantiation and spawn time breakdown on OpenStack and Nomad}
  \label{fig:vims_inst_time} 
  \vspace{-3ex}
\end{figure*}

Two main results are hereafter presented: i) the total time needed to instantiate a ClickOS VM (as an example of a Micro-VNF); ii) the timing breakdown of the spawning process in OpenVIM, OpenStack and Nomad. Since the results of the \textit{tuned} OpenVIM were very promising we have better examined the performance analysis for this VIM. Compared to our previous work, we have introduced performance evaluation for the OpenStack and OpenVIM virtual infrastructure managers and omitted the results related to OpenStack Legacy: performance evaluation of OpenStack with Neutron is of higher relevance since this represents the long-term supported configuration. Moreover, for the sake of clarity we prefer to maintain a three-way comparison. Major details about the OpenStack Legacy evaluation can be found in \cite{piervim}.

The first results that we report are based on the VIM Message Analyzer that we have developed~\cite{ourgitrepo} (except for OpenVIM, as described above) and presented in Figure~\ref{fig:vims_inst_time}.
The timing breakdown of the spawning process has been obtained with the approach of inserting timestamp loggers into the source code of the VIMs under evaluation. All results were obtained by executing a test of 100 replicated runs, in unloaded conditions. Error bars in the figures denote the 95\% confidence intervals of the results. In each run we requested the instantiation of a new ClickOS Micro-VNF. The VM was deleted before the start of the next run. As regards OpenVIM, we analyze its performance also under loaded conditions and submitting batches of requests.

Our experimental setup comprises two hosts with an Intel Xeon 3.40GHz quad-core CPU and 16GB of RAM. One host (hereafter referred to as HostC) is used for the VIM core components, and the other host (HostL) for the VIM local components and the compute resource. We use Debian 8.3 with Xen-enabled v3.16.7 Linux kernels. Both hosts are equipped with two Intel X540 NICs at 10 Gb/s where: one interface has been used as the management interface and the other one for the direct interconnection of the host (data plane network). In order to emulate a third host running OpenStack Rally, Nomad Pusher and OpenVIM Client CLI, we created a separate network namespace using the \textit{iproute2} suite in HostC, then we interconnected this namespace to the data plane network. This approach is useful also for the \textit{VIM Message Analyzer} because it offers a practical way to better understand the exchange of messages during the different phases (since they come from different sources).

\subsection{OpenStack Experimental Results}
\label{sec:openstackresults}

For the OpenStack setup, we run Keystone, Glance, Nova orchestrator, Horizon, Neutron server and the other components in HostC, while we run Nova Compute and Neutron Agents in HostL. We use the development version of OpenStack (DevStack~\cite{openstack}) for our deployment, which is more suitable for our experimental testbed, since it quickly brings up a complete OpenStack environment based on specific releases or on modified versions and it allows to incrementally bring up and down OpenStack services when needed. We have adapted it to run on the two available servers because typically it envisages a single machine deployment. We use the modifications presented in Section~\ref{sec:openstackmod} and the \textit{libvirt/libxl} support in OpenStack in order to boot ClickOS Xen machines.

With reference to the instantiation model depicted in Figure~\ref{fig:vimNovaNeutron}, we report in Figure~\ref{fig:osneutronclickos} the measurements of the instantiation process in OpenStack, separated by component. The upper horizontal bar (\textit{Stock}) refers to the OpenStack version that only includes the modifications needed to boot the ClickOS VMs (as described in Section~\ref{sec:openstackmod}). Note that in the tested deployment we assume that the Block Storage service (Cinder) has already created all the needed block devices. This implies that the block device creation time will not be considered in our test, while we count its retrieval time. The experiment reports a total time exceeding two seconds for the \textit{stock} version, which is not adequate for highly dynamic NFV scenarios. Analyzing the timing of the execution of the single components, we note that almost one second is spent during the spawning phase while the sum of other components accounts for around 1.1 seconds.

In Figure~\ref{fig:neutronclickos} (upper horizontal bar), we show the details of the spawning phase which is split into three sub-phases: 1) \textit{Create image}, 2) \textit{Generate XML} and 3) \textit{Libvirt spawn}. The first two sub-phases are executed by the Nova libvirt driver, and the last one is executed by the underlying libvirt layer. By analyzing the timing of the \emph{stock} version, we can see that the \textit{Create image} is the slowest step with a duration of about 0.5 seconds. During the \textit{Create image} phase, the Nova libvirt driver waits for Cinder to retrieve of the block device and for Neutron to create/retrieve the network ports and connections. The image creation step includes operations such as the creation of log files and of the folders to store Glance images. The original OS image (the one retrieved from Glance) is re-sized in order to meet user requirements (the so called \textit{flavors} in OpenStack's jargon). Moreover, if it is required, swap memory or ephemeral storage are created, and finally, some configuration parameters are injected into the image (e.g., SSH key-pair, network interface configuration). The \textit{Generate XML} and \textit{Libvirt spawn} steps introduce a total delay of about 0.4 seconds. During the \textit{Generate XML} step, the configuration options of the guest domain are retrieved and then used to build the guest domain description (the XML file given as input to \textit{libvirt}). For instance, options such as the number of CPUs and CPU pinning are inserted in the XML file. Once this step is over, the libxl API is invoked in order to boot the VM. When the API returns, the VM is considered spawned, terminating the instantiation process. In this test, we are not considering the whole boot time of the VM, as this is independent of the VIM operations, and thus out of the scope of this work. The considered spawning time measures only the time needed to create the Xen guest domain.

The performance measurements for OpenStack reported above show that most of the instantiation time is spent during the spawning phase and in particular on the \textit{Create image} operation. In this work and also in the previous one, we did not target the optimization of other components: OpenStack being a general purposes VIM, tailoring and optimizing it for a single scope (e.g., managing only the VNF lifecycle) would completely distort its purpose. Moreover, some components are already optimized, for example it was not possible to implement meaningful optimizations on Neutron, as it is already tuned and makes extensively use of caching. Caches are also employed in other operations: image download, block device creation/reuse, network creation. However, it is still possible to tune the spawning phase without losing the general scope of OpenStack.

The \textit{tuned} horizontal bar in Figure~\ref{fig:osneutronclickos} and Figure~\ref{fig:neutronclickos} report the obtained result. The optimizations did not alter too much the workflow shown in Figure~\ref{fig:vimNovaNeutron} as they only exclude Cinder from the process and they make smaller the \textit{Spawning} in the Nova \textit{libvirt} driver. The reduction of the instantiation time highlights that shorter boot times are not the only advantage of Unikernels compared to full fledged OSs: specific VIM optimizations can be performed due to their characteristics. We are using a tiny diskless VM, which means we can skip most of the actions performed during the image creation step. For example, we do not need to resize the image in order to match the flavor disk size. Moreover, Unikernels provide only the functionality needed to implement a specific VNF, so it is unlikely to have an SSH server running on it, hence SSH key-pairs configuration is not needed. The tuned version does not include the operations related to block device mapping, due to ClickOS machines being diskless. After this optimization, the \textit{Create image} phase has been ``shortened''. However, it still has an impact of 50ms in the tuned boot. The removal of the block device operations also has an effect on the spawning phase, removing the need to wait for block device information.  The \textit{Generate XML} phase is also affected, as fewer operations have to be performed. Implementing these optimizations, we are able to reduce the spawning time down to 0.32 s (Figure~\ref{fig:neutronclickos}, bottom bar) obtaining a reduction of about 65\% compared to the stock version of OpenStack. Looking at the overall instantiation time, the relative reduction (Figure~\ref{fig:osneutronclickos}, bottom bar) is about 30\%, down to 1.5 s from 2.1 s for the \textit{stock} OpenStack.

\subsection{Nomad Experimental Results}
\label{sec:nomadresults}

According to the Nomad architecture, the minimal deployment comprises two nodes. Therefore, we have deployed a Nomad Server, which performs the scheduling tasks, on HostC and a Nomad Client, which is responsible for the instantiation of virtual machines, on HostL.

In Section~\ref{sec:modelling}, we have identified two major steps in the breakdown of the instantiation process: scheduling and instantiation. The upper horizontal bar in Figure~\ref{fig:nomad_1_clickos} reports the results of the performance evaluation for the \textit{stock} Nomad. The total instantiation time is much lower than 1 sec. This result is not surprising: as discussed in Section~\ref{sec:background}, Nomad is a minimalistic VIM providing only what is strictly needed to schedule the execution of virtual resources and to instantiate them. Looking at the details, the scheduling process is very light-weight, with a total run time of about 50 ms. The biggest component in the instantiation time is the spawning process which is executed by the XenDriver.  Diving into the driver operations, we identified 4 major steps: \textit{Download artifact}, \textit{Init Environment}, \textit{Spawn}, and \textit{Clean}, as reported in in Figure~\ref{fig:nomad_2_clickos}. In the first step, Nomad tries to download the artifacts specified by the job. For a Xen job, the client is required to download the configuration file describing the guest and the image to load. This operation adds a delay of about 40 ms and can be optimized or entirely skipped. \textit{Init Environment} and \textit{Clean} introduce a low delay (around 20 ms) and are interrelated: the former initializes data structures, creates log files and folders for the \textit{Command executor}, the latter cleans up the data structures once the command execution is finished. The \textit{XL spawn} is the step which takes longer but by studying the source code we found no room to implement further optimizations: indeed the total spawning measured time is around 100 ms. Considering a light overhead introduced by the \textit{Command executor} the time is very similar to the what we obtain by running directly the \textit{xl} toolstack. The overall duration of the spawning phase is 160 ms, lower that the 280 ms for the instantiation phase reported in Figure~\ref{fig:nomad_1_clickos}. This is due to the notification mechanism from the client towards the server. It uses a lazy approach for communicating the end of the scheduled jobs: when the message is ready to be sent, the client waits for a timer expiration to attempt to aggregate more notifications in a single message. This means that the instantiation time with Nomad is actually shorter than the one shown in Figure~\ref{fig:nomad_1_clickos}.

Thus, we have observed that the instantiation times are already smaller, in the order of 0.3 s (or 0.2 s if we do not consider the notification delay from Nomad Client to Nomad server). This can be explained by the following: i) the scheduler in the Nomad Server is very lightweight and its functionality is much simpler; ii) in the client we have implemented from scratch a new Nomad driver for Xen (cf.\ Section~\ref{sec:modifications}) and hence included only code in the driver which is strictly necessary to interact with \textit{XL}. Starting from our initial implementation, we introduced further improvements streamlining the execution of the Nomad Client (the \textit{Download Artifact} step of the XenDriver) assuming that the images of the Micro-VNFs can be stored locally. Pursuing this approach we can reduce the Driver operation by about 30 ms (see Figure~\ref{fig:nomad_2_clickos}).

\subsection{OpenVIM Experimental Results}
\label{sec:openvimresults}

We deploy \textit{OpenVIM Core} components and \textit{OpenVIM DB} on HostC, while HostL has been employed as compute node. We use the \textit{development} mode for OpenVIM in our experiments, which requires less resources and disables some functionalities, like EPA features, passthrough and many others. We make use of the modifications presented in Section~\ref{sec:openvimmod} in order to boot ClickOS Xen machines. We have also configured HostC as compute node noticing no performance degradation of the system.

\begin{figure}
     \centering
     \includegraphics[width=0.485\textwidth]{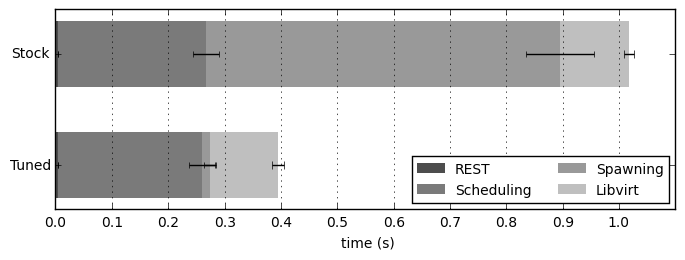}
     \caption{ClickOS instantiation time on OpenVIM}
     \label{fig:ovimclickos}
     \vspace{-3ex}
\end{figure}

\begin{figure*}[!t] 
    \centering
  \subfloat[OpenVIM stock]{%
       \includegraphics[width=0.485\textwidth]{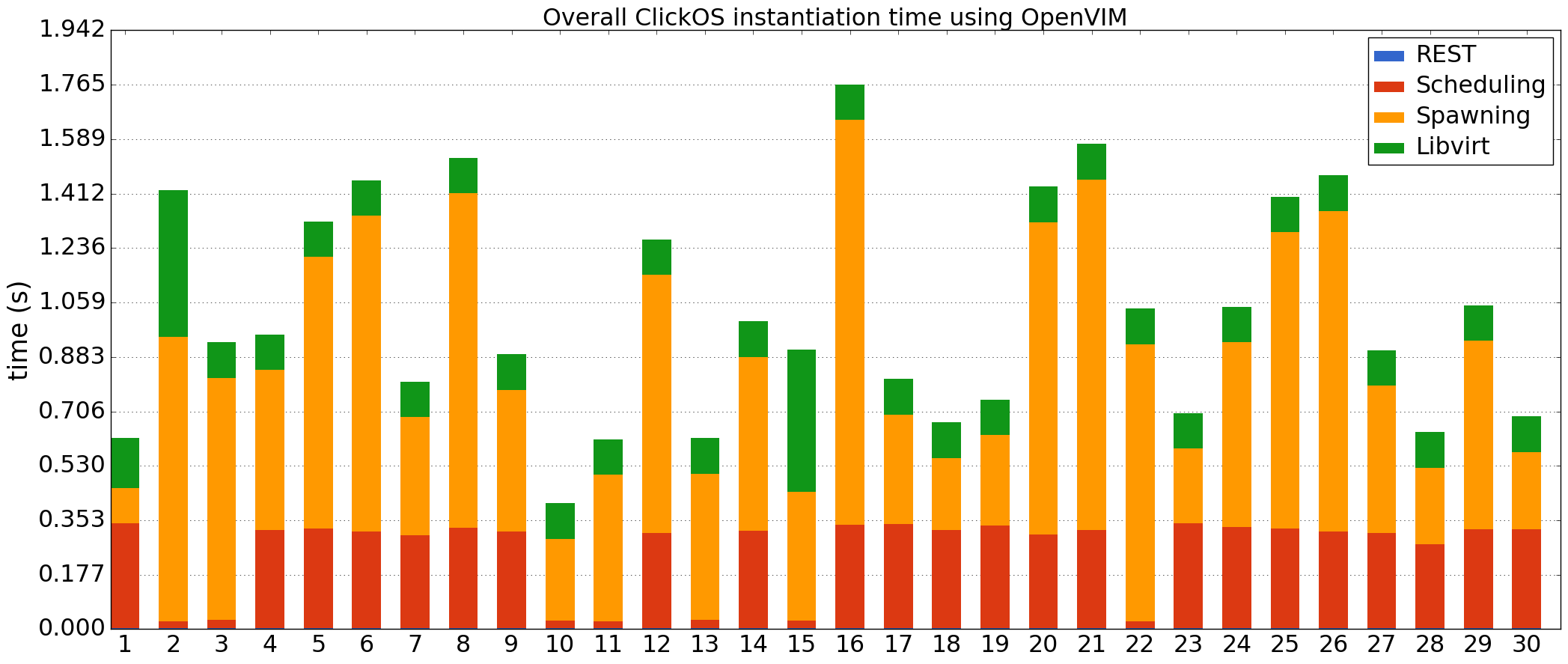}
    \label{fig:openvimclickos}}\hfill
  \subfloat[OpenVIM tuned]{%
  \includegraphics[width=0.485\textwidth]{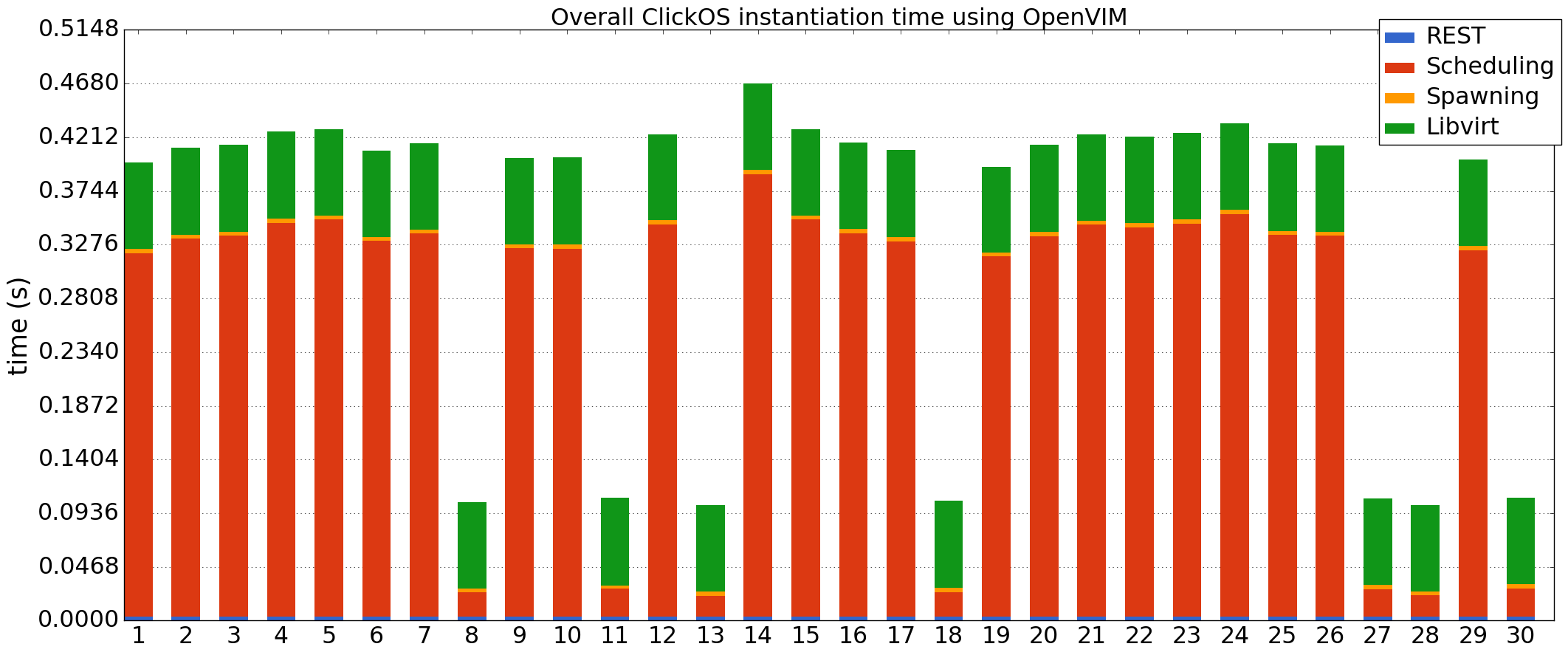}
    \label{fig:openvimclickostuned}}
  \vspace{-2ex}
  \caption{Instability analysis of OpenVIM}
  \vspace{-3ex}
\end{figure*}

The upper horizontal bar in Figure~\ref{fig:ovimclickos} shows the measurements related to the instantiation process for OpenVIM \textit{stock} divided in \textit{REST}, \textit{Scheduling}, \textit{Spawning} and \textit{libvirt} phases. With reference to the model depicted in Figure~\ref{fig:vimOpenVIM}, \textit{REST}, \textit{Scheduling} and part of the \textit{Spawning} are mapped to the \textit{OpenVIM Core} and \textit{MySQL} components. The second part of the \textit{Spawning} and the \textit{libvirt} phases are spent on the compute node and they are mapped respectively with \textit{OpenVIM Local} and \textit{libvirt} components. Summing up all the parts we have obtained a total instantiation time of about 1 s.

Our analysis of this behavior is reported later in this section. We have started the optimization process by excluding from the instantiation phase all the operations which are not required for the ClickOS boot but performed in any case by OpenVIM. With reference to the instantiation workflow shown in Figure \ref{fig:vimOpenVIM}, our tuning mainly affects the duration of the \textit{Spawning} phase in the \textit{OpenVIM Core} and \textit{OpenVIM Local} components. During the \textit{Spawning} process, the VM interface is reset sending \textit{ifdown/ifup} via SSH. This operation is totally irrelevant for ClickOS and it can be very time consuming since the \textit{Spawning} is blocked waiting for the interfaces coming down/up. Another operation we have optimized, it is the full-copy of the VM image: by default a copy is performed, inside the compute node, moving the VM image from the OpenVIM folder in the \textit{libvirt} folder. We have removed this overhead, maintaining a local copy of the image in the \textit{libvirt} folder and thus saving a \textit{cp} command each time. \textit{libvirt} is a source of overhead itself: by default operates using SSH as transport layer. Even if all its operations are small, SSH can introduce a certain amount of overhead. Thus, as further optimization, we have enabled TCP as transport layer for the \textit{libvirt} communication channel instead of SSH. Finally, we have optimized the generation of the XML (like we have done for OpenStack) because for the \textit{stock} version we have used the default behavior of \textit{libvirt} with Xen. We have removed all unnecessary configuration parts like the graphics, serial ports and we have avoided also the allocation of the CD-ROM virtual device and of the hard disk. Then, we have focused our attention on the \textit{OpenVIM Core} components. In particular, the management of work queues has attracted our attention: for each compute node there is a worker thread which manages a working queue where the instantiation requests are enqueued as output of the scheduling. By default, the worker checks the queue each second. In a deployment with a few numbers of compute nodes, this obviously is too conservative. Setting this refresh time to zero has led to a considerable reduction of the instantiation process, but at the same time has increased the load on the CPU. With our deployment we have found a good compromise setting up the polling interval at 1 ms (in this configuration, we have measured a load of 2\% in the HostC where \textit{OpenVIM Core} and \textit{OpenVIM DB} run). Results of our optimizations are shown in bottom row of the Figure~\ref{fig:ovimclickos}. With the above modifications we are able to measure times of around 400 msec. Looking at the overall instantiation time, the relative reduction is about 65\%. The \textit{Spawning} phase becomes negligible and stable (it is possible appreciate an high variance in the stock version), while the other phases are mostly unchanged. As regards \textit{libvirt} this is expected, while the results of the \textit{Scheduling} phase were quite suspect.

\begin{figure*}[!t] 
    \centering
  \subfloat[Instantiation time breakdown on OpenVIM tuned]{%
       \includegraphics[width=0.485\textwidth]{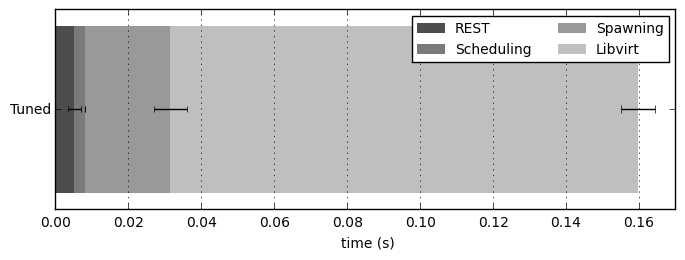}
    \label{fig:ovimclickos2}}\hfill
  \subfloat[Spawning time breakdown on OpenVIM tuned]{%
  \includegraphics[width=0.485\textwidth]{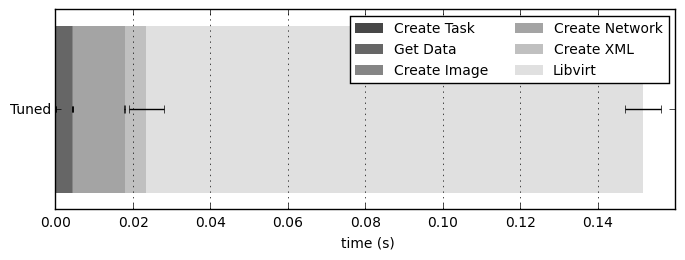}
    \label{fig:openvimclickos2}}
  \caption{ClickOS instantiation and spawn time breakdown on OpenVIM}
  \vspace{-3ex}
\end{figure*}

Looking at Figure~\ref{fig:openvimclickostuned}, coefficient of variation is around 40\% for the \textit{Scheduling} phase, which is high if compared to the other VIMs (also in OpenStack with an improved scheduling we did not appreciate this high variance). We have dug into the results of the single runs and we have noticed strange behavior. Figure~\ref{fig:openvimclickos} reports this analysis showing the results of the first 30 runs of the \textit{stock} version, while Figure~\ref{fig:openvimclickostuned} shows the same for the \textit{tuned} version of OpenVIM. It is possible to note that in the \textit{tuned} version the variance related to \textit{Spawning} phase completely disappeared after our optimizations. While, it is obvious that there is high irregularity in the runtime of the \textit{Scheduling} part. We have analyzed the internals of this part and we have noticed that is composed by a decisional component and by the storing of the decision in the database. In order to analyze it further, we have inserted additional timestamps to split the decision of the scheduler from the storing of the decision into the MySQL database. The separation of the times has shown a decision phase with an average of 2.21 ms, while the operation of storing into the database has presented strange results with a bistable behavior. At this point we have repeated the same tests enabling the profiling of the code in OpenVIM. We have identified that the high variability of execution time was related to the \textit{commit} function of the \textit{\_msql.connection} object. In order to solve the problems we were facing, first we have tried different configuration for MySQL with no success. We made the hypothesis that we had problems with the hard disk and recreated a small testbed using two Intel NUC personal computers equipped with Intel Core I5 processors, 16GB of RAM and 1 Gb/s Ethernet. We repeated the experiments and the instability problems related to the \textit{OpenVIM DB} disappeared, this confirmed our hypothesis that the problem was in the hard drive of our machines. Leveraging the fact the deployment of OpenVIM is not resource-hungry with regards to disk space, we have changed the setup of OpenVIM deploying all the components on \textit{ramdisk}. Figure~\ref{fig:ovimclickos2} shows the results of this operation. In Figure~\ref{fig:openvimclickos2} we have reported also a more detailed breakdown. The resulting times are very low and demonstrate that we have circumvented the problems related to the database. With the setup described above, we measure times around 0.15 seconds. Looking at the overall instantiation time, the relative reduction is about 57\% with respect to the results of the first \textit{tuned} version and 85\% with respect to the \textit{stock} version of OpenVIM.

It can be noted from the results that it was not possible to reduce the component brought by \textit{libvirt}. This time is mainly composed of: i) the system calls done to spawn the Unikernel; ii) the setup of the network; iii) the interaction with the \textit{xenstore}. In particular for what concern the setup of the network, we measure a time of 31 msec if we use Open vSwitch as bridge and 14 msec if we use Linux Bridge. We obtain these times emulating the network configuration: \textit{libvirt}, during this phase, gives the control to Xen, which uses the scripts in the folder \textit{/etc/xen/scripts} to setup the network, starting from these we have emulated Unikernel instantiations and we have measured the network overhead. In the light of this, the entire remaining part is spent by the \textit{libvirt} system calls and by the \textit{xenstore}. The former and the networking part, although it consists only on connecting the VM with its bridge, are still required. While, the second part can be further optimized as shown in \cite{manco2017my} where authors propose a redesigned toolstack for Xen without the \textit{xenstore} component.

Since the results are very promising, we have extended our performance evaluation of OpenVIM with two additional benchmarks. In the first one we have tested OpenVIM with different load conditions and we have recorded the total instantiation times. Table~\ref{tab:ovimclickosstress} reports the results of this experiment. To generate different loads, we instantiate a number N of VMs as background, then we execute a number of runs where we recorded the time to instantiate the N+1 VM. In the table, the first column (namely Load) reports exactly the number of VMs already spawned, while we show the times generated by the instantiation of an additional VM in the other columns. In order to spawn a high number of VMs we have pinned one core to Dom0 (two CPUs), scheduling the VMs on the remaining cores.  Otherwise, the risk is to have CPU stuck during the experiments. The results show a monotone increase with the number of VMs. It is possible to appreciate that \textit{Spawning} stays stable with almost negligible increases, \textit{REST} and \textit{Scheduling} show a very slow tendency to increase. Instead, \textit{libvirt} time results to be influenced by the number VMs, with times that start around 0.15 seconds when the system is unload to arrive to 0.4 seconds with 300 VMs running in background. This matches \cite{manco2017my}, where the authors provided a deep study of the instantiation overhead, and pointed out that Xen's \textit{xenstore} leads to a super-linear increase of instantiation times as the number of VMs on a system increases.

In the last experiment, we have submitted batch requests with the objective of simulating a VNFs chain instantiation. Each batch contains the instantiation request of a number VMs M which varies between 5 and 10. OpenVIM performance has been evaluated considering different initial states for the system (0, 10, 50, 100, 200 VMs already running). We did not include results for 300 VMs since we were not able to finish the experiments without errors. Using this hardware we were able to spawn a number of VMs slightly higher than 300, this had been verified using only \textit{xl} toolstack. 

 \begin{table}[ht]
    \scalebox{1}{
	    \begin{tabular}{|c|c|c|c|c|}\hline
        \multicolumn{5}{|c|}{\textbf{Instantiation time breakdown (s)}}\\\hline
	      \textbf{Load} & \textbf{REST} & \textbf{Scheduling}  & \textbf{Spawning} & \textbf{libvirt} \\
	      \hline
	      10 VM  &  $0.004$ & $0.004$ & $0.185 \pm 0.043$ & $0.157 \pm 0.004$ \\
	      \hline
	      20 VM  &  $0.007$ & $0.004$ & $0.216 \pm 0.035$ & $0.161 \pm 0.003$ \\
	      \hline
	      50 VM  &  $0.007$ & $0.004$ & $0.192 \pm 0.021$ & $0.181 \pm 0.005$ \\
	      \hline
	      100 VM & $0.007$ & $0.004$ & $0.205 \pm 0.016$ & $0.216 \pm 0.008$ \\
	      \hline
	      150 VM & $0.007$ & $0.004$ & $0.207 \pm 0.013$ & $0.254 \pm 0.011$ \\
	      \hline
	      200 VM & $0.007$ & $0.005$ & $0.206 \pm 0.011$ & $0.297 \pm 0.012$ \\
	      \hline
	      250 VM & $0.007$ & $0.005$ & $0.195 \pm 0.012$ & $0.346 \pm 0.015$ \\
	      \hline
	      300 VM & $0.008$ & $0.006$ & $0.195 \pm 0.011$ & $0.401 \pm 0.018$ \\
	      \hline
	    \end{tabular}
	    }
    \centering \caption{OpenVIM under load conditions}
    \label{tab:ovimclickosstress}
    \vspace{-2ex}
 \end{table}

Table~\ref{tab:ovimclickosbatch} reports the results of this experiment. Values are obtained averaging the instantiation time of the VMs booted during the run, in each run a batch of size M is submitted. The results are higher than the previous tests, but this is expected since we are submitting several jobs at this same time, which introduces overhead with respect to the instantiation of a single VNF. In our configuration we pinned one core to Dom0, thus spawning more VMs at the same time will result in more contention of the two CPUs we provided to Dom0. Analyzing the results, it is possible appreciate that these higher values are mainly due to \textit{libvirt} while other components are stable. Moreover, passing from unloaded conditions to 10 VMs already instantiated introduces an overhead about 100 msec in the database reads, which remains constant in the tests with 50 VMs, 100 VMs and 200 VMs. Instead, changing the size of the batch in the considered values does not introduce a considerable overhead.

   \begin{table}[ht]
    \scalebox{1}{
	    \begin{tabular}{|c|c|c|c|c|c|c|}\hline
        \multicolumn{7}{|c|}{\textbf{Instantiation time (s)}}\\\hline
	      \textbf{Load} & \textbf{5 VM} & \textbf{6VM} & \textbf{7VM} & \textbf{8 VM} & \textbf{9 VM} & \textbf{10 VM} \\
	      \hline
	      0 VM   & $0.319$ & $0.337$ & $0.342$ & $0.371$ & $0.374$ & $0.391$ \\
	      \hline
	      10 VM  & $0.447$ & $0.467$ & $0.461$ & $0.455$ & $0.454$ & $0.463$ \\
	      \hline
	      50 VM  & $0.535$ & $0.539$ & $0.526$ & $0.545$ & $0.551$ & $0.543$ \\
	      \hline
	      100 VM & $0.622$ & $0.594$ & $0.624$ & $0.616$ & $0.622$ & $0.632$\\
	      \hline
	      200 VM & $0.814$ & $0.829$ & $0.841$ & $0.862$ & $0.851$ & $0.863$\\
	      \hline
	    \end{tabular}
	    }
    \centering \caption{OpenVIM under different load conditions and batch submissions}
    \label{tab:ovimclickosbatch}
    \vspace{-2ex}
 \end{table}

\subsection{Gap Analysis and Discussion of the Results}
\label{sec:discussion}

Considering the results of the performance evaluation we believe that a step towards the support of highly dynamic scenarios has been done but there is still an important performance gap to be filled in the orchestrators to fully support the Superfluid NFV scenarios. The tuned version of OpenStack shows an improvement of about 30\% but the absolute value of the instantiation times is still around 1.5 seconds which is too high for the considered scenarios. This would result in an additional delay for the first packet coming from the user when a Unikernel is spawned at the time of arriving of the packet. As regards Nomad, the instantiation times are small at least when the system is unloaded, however Nomad is a very minimalistic orchestrator and lacks of useful functionality (like the support for advanced networking features) that we can found in OpenStack and OpenVIM. As regards OpenVIM, it showed good instantiation times (around 0.15 seconds) and a good base of functionality. However, the experiments executed under loaded conditions exhibited unsatisfactory performance with a super-linear increase of instantiation times as the system become loaded.

We learned some lessons from this work, which could be applied to the design of new VIMs or to the refactoring of existing VIMs. Modularity is an important characteristic to turn the VIMs into composable projects and adapt them to new requirement. Initially, we felt lost and unable to do also small modifications without resorting to hacks. VIMs should not force the users to a predefined workflow but it should be possible to exclude some steps without explicitly killing the components of the VIMs. We advocate that new VIMs or existing VIMs should be more modular to accommodate new features more easily, like the support for other hypervisors or new tools (for example in \cite{manco2017my}, the authors propose an optimized toolstack for Xen). Another consideration is that Unikernels introduce an important shift from general purpose VMs to specialized VMs: full VMs can be easily configured after they are deployed, while for Unikernels like ClickOS most reconfigurations need to happen in the image creation phase, before that the Unikernel is deployed. VIMs have to take into account this and the toolchain of the Unikernels should be integrated in the workflow of the VIMs, easing the deployment of customized Unikernels.

\vspace{-1ex}
\section{Related Work}
\label{sec:relatedwork}

VIM performance evaluation is a topic addressed also by other works. In~\cite{OpenStackCloudStack,OpenStackvsEucalyptus,OpenStackOpenNebula} authors compare the performance of OpenStack versus other projects (CloudStack, OpenNebula, Eucaliptus). However, the performance figures of VIMs are analyzed in terms of time needed to instantiate fully fledged VMs and they are mainly focused on mere benchmarking without a deep analysis of the tasks performed during the instantiation. On the other hand, in~\cite{callegati2014performance,litvinski2013experimental}, authors consider only OpenStack and focus on particular aspects such as networking components rather than the scheduling. Other works (such as~\cite{martins2014clickos, manco2014towards, manco2015case}) focus on the performance of ClickOS and of the NFVI. They demonstrate that it is possible to guarantee low latency instantiation times for Micro-VNFs and the suitability of ClickOS for the NFV and Superfluid Cloud use cases. Instead, our work is focused on the analysis of the performance of VIMs and of their suitability in the NFV frameworks (including also the Superfluid NFV framework). 

In~\cite{ledyayev2014high} the authors consider performance tuning of OpenStack-based virtual infrastructures, targeting the high performance computing scenario. Instead, our work is aimed at the NFV scenario and considers a broader set of virtual infrastructure technologies. Enos, described in~\cite{cherrueau2017toward}, is an open source toolset to automatize the deployment and conduct experiments on OpenStack. The tools that we use in our work are released as open source as well, but aim at a plurality of VIMs.

\cite{ieeeComms16rm, ieeeNFVsurvey} describe solutions covering the whole NFV framework and for the implementation of ETSI MANO specifications.
Among these, solutions such as CloudBand, CloudNFV and OpenNFV are proprietary and there are not enough details available to include them in our analysis, while OPNFV defines a general framework for NFV, and, as for the VIM component, it employs OpenStack. However these works do not address the performance of the practical implementations of the NFV framework. TeNOR is another open source project which implements the NFV MANO framework. 
In \cite{tenor}, the authors report a performance evaluation which addresses only TeNOR at a high level, while our work aims at a broader and deeper scope.

This work significantly extends the preliminary results described in~\cite{piervim}: i) the analysis and the modelling of the instantiation process consider also OpenVIM from the OSM project~\cite{etsiosm} and OpenStack with Neutron network component (replacing Nova network, which is now deprecated); ii) A new modification has been introduced for OpenStack that allows to deploy Unikernels images without resorting any hack or changing the execution of the VIM; iii) OpenVIM has been modified as well in order to integrate Xen hypervisor and instantiate Unikernels based on ClickOS; new modifications have been implemented for tuning; iv) performance evaluation tools have been extended to support new VIMs under analysis; v) Analysis and experimental work consider also OpenVIM and introduce new results for OpenStack (considering its latest release); vi) finally, since OpenVIM results were very promising we have evaluated OpenVIM performances under load conditions and submitting batch of requests with the objective of emulating different workloads including also the instantiation of VNFs chains.

\vspace{-1ex}
\section{Conclusions}
\label{sec:conclusions}

In this work, we have described a general reference model of the VNF and Micro-VNF instantiation process, and we have used it to describe the internal workflow of three VIMs: OpenStack with Nova networking, OpenStack with Neutron networking, OpenVIM and Nomad. We have considered the instantiation of ClickOS as an example for a Micro-VNF, which enables highly dynamic scenarios like the Superfluid NFV Clouds that have been the starting point of this study. 

A critical step of our work has been the implementation of the modifications needed to integrate the Xen hypervisor and the instantiation of Unikernels based on ClickOS in the aforementioned VIMs. Leveraging these, we have provided measurements on the instantiation process in an experimental setup, using ad-hoc developed profiling tools. The modifications of the VIMs and the tools we have developed are available as open source \cite{ourgitrepo, etsiosm}. Starting from the analysis of the measurements results, we have also optimized the instantiation times of Micro-VNFs by modifying the source code of the VIMs. \optional{For the performance measurements we have considered the basic case of an unloaded system and of a very simple physical deployment. We have measured the instantiation time of a VM considering a single request (no other background requests in parallel), assuming that the requested VM is the first one to be allocated/scheduled (the database of allocated VMs is empty) and that only one target node is available for deploying the VM. Since OpenVIM results were very promising, we have extended our performance evaluation with two additional benchmarks targeting more realistic workloads.} To the best of our knowledge there are not results evaluating the suitability of the state of the art technologies for highly dynamic scenarios nor works doing a performance evaluation of the NFV framework with same breadth and comparing several VIMs.

Potential directions of work are to further improve the performance of the considered VIMs. Currently, we can identify two viable approaches: i) optimization of other components; ii) pre-instantiation of Micro-VNFs. We want to highlight that many components have not been considered in this work during the optimization steps (e.g., trying to replace the lazy notification mechanism of Nomad with a reactive approach). It would also be interesting evaluate the feasibility of scenarios where the Micro-VNFs are preallocated in batches and left in pause as long as a new requests are submitted. This has been investigated in \cite{manco2017my}, but only on a hypervisor, with no orchestration support.

\vspace{-2ex}
\section*{Acknowledgment}
This paper has received funding from the EU H2020 Superfluidity and 5G-EVE projects.



%

\vspace{-1ex}
\bibliographystyle{IEEEtran}
\bibliography{short_bib}

\begin{thebibliography}{10}
\providecommand{\url}[1]{#1}
\csname url@samestyle\endcsname
\providecommand{\newblock}{\relax}
\providecommand{\bibinfo}[2]{#2}
\providecommand{\BIBentrySTDinterwordspacing}{\spaceskip=0pt\relax}
\providecommand{\BIBentryALTinterwordstretchfactor}{4}
\providecommand{\BIBentryALTinterwordspacing}{\spaceskip=\fontdimen2\font plus
\BIBentryALTinterwordstretchfactor\fontdimen3\font minus
  \fontdimen4\font\relax}
\providecommand{\BIBforeignlanguage}[2]{{%
\expandafter\ifx\csname l@#1\endcsname\relax
\typeout{** WARNING: IEEEtran.bst: No hyphenation pattern has been}%
\typeout{** loaded for the language `#1'. Using the pattern for}%
\typeout{** the default language instead.}%
\else
\language=\csname l@#1\endcsname
\fi
#2}}
\providecommand{\BIBdecl}{\relax}
\BIBdecl

\bibitem{nfv}
\BIBentryALTinterwordspacing
``{ETSI Network Function Virtualization}.'' [Online]. Available:
  \url{http://www.etsi.org/technologies-clusters/technologies/nfv}
\BIBentrySTDinterwordspacing

\bibitem{cloud}
R.~Buyya \emph{et~al.}, ``{Cloud computing and emerging IT platforms: Vision,
  hype, and reality for delivering computing as the 5th utility},''
  \emph{Future Generation computer systems}, vol.~25, no.~6, 2009.

\bibitem{huicicomparison}
\BIBentryALTinterwordspacing
F.~Huici \emph{et~al.}, ``{VMs, Unikernels and Containers: Experiences on the
  Performance of Virtualization Technologies}.'' [Online]. Available:
  \url{https://www.ietf.org/proceedings/95/slides/slides-95-nfvrg-2.pdf}
\BIBentrySTDinterwordspacing

\bibitem{manco2017my}
F.~Manco \emph{et~al.}, ``{My VM is Lighter (and Safer) than your Container},''
  in \emph{26th Symposium on Operating Systems Principles}.\hskip 1em plus
  0.5em minus 0.4em\relax ACM, 2017, pp. 218--233.

\bibitem{shetty2017empirical}
J.~Shetty \emph{et~al.}, ``{An Empirical Performance Evaluation of Docker
  Container, OpenStack Virtual Machine and Bare Metal Server},''
  \emph{Indonesian Journal of Electrical Engineering and Computer Science},
  vol.~7, no.~1, pp. 205--213, 2017.

\bibitem{xenwh}
{Xen Project}, ``{The Next Generation Cloud: The Rise of the Unikernel},''
  Linux Foundation, Tech. Rep., 2015.

\bibitem{cormack}
\BIBentryALTinterwordspacing
{J. Cormack}, ``{The Modern Operating System in 2018}.'' [Online]. Available:
  \url{https://www.youtube.com/watch?v=dR2FH8z7L04}
\BIBentrySTDinterwordspacing

\bibitem{unikraft}
\BIBentryALTinterwordspacing
``{Unikraft Project}.'' [Online]. Available:
  \url{https://www.xenproject.org/linux-foundation/80-developers/207-unikraft.html}
\BIBentrySTDinterwordspacing

\bibitem{clearwater}
\BIBentryALTinterwordspacing
``{Project Clearwater}.'' [Online]. Available:
  \url{http://www.projectclearwater.org}
\BIBentrySTDinterwordspacing

\bibitem{cord}
\BIBentryALTinterwordspacing
``{OpenCORD Project}.'' [Online]. Available: \url{https://opencord.org}
\BIBentrySTDinterwordspacing

\bibitem{manco2014towards}
F.~Manco \emph{et~al.}, ``{Towards the Superfluid Cloud},'' in \emph{ACM
  SIGCOMM Computer Comm. Rev.}, vol.~44.\hskip 1em plus 0.5em minus 0.4em\relax
  ACM, 2014, pp. 355--356.

\bibitem{manco2015case}
F.~Manco \emph{et~al.}, ``{The Case for the Superfluid Cloud},'' in
  \emph{HotCloud}, 2015.

\bibitem{SF-architecture}
G.~Bianchi \emph{et~al.}, ``{Superfluidity: a flexible functional architecture
  for 5G networks},'' \emph{Trans. on Emerging Telecommunication Technologies},
  vol.~27, no.~9, 2016.

\bibitem{martins2014clickos}
J.~Martins \emph{et~al.}, ``{ClickOS and the art of Network Function
  Virtualization},'' in \emph{11th USENIX Conference on Networked Systems
  Design and Implementation}.\hskip 1em plus 0.5em minus 0.4em\relax USENIX
  Association, 2014.

\bibitem{piervim}
P.~L. Ventre \emph{et~al.}, ``{Performance Evaluation and Tuning of Virtual
  Infrastructure Managers for (Micro) Virtual Network Functions},'' in
  \emph{Network Function Virtualization and Software Defined Networks
  (NFV-SDN), IEEE Conference on}.\hskip 1em plus 0.5em minus 0.4em\relax IEEE,
  2016, pp. 141--147.

\bibitem{ourgitrepo}
\BIBentryALTinterwordspacing
``{VIM} tuning and evaluation tools.'' [Online]. Available:
  \url{https://github.com/netgroup/vim-tuning-and-eval-tools}
\BIBentrySTDinterwordspacing

\bibitem{etsiosm}
\BIBentryALTinterwordspacing
``{O}pen {S}ource {M}ano.'' [Online]. Available: \url{https://osm.etsi.org/}
\BIBentrySTDinterwordspacing

\bibitem{nomad}
\BIBentryALTinterwordspacing
``{Nomad Project}.'' [Online]. Available: \url{https://www.nomadproject.io}
\BIBentrySTDinterwordspacing

\bibitem{openstack}
\BIBentryALTinterwordspacing
``{O}pen{S}tack.'' [Online]. Available: \url{https://www.openstack.org}
\BIBentrySTDinterwordspacing

\bibitem{mano}
{\relax ETSI}.~{G}roup~for {NFV}, ``{N}etwork {F}unctions {V}irtualisation
  ({NFV}); {M}anagement and {O}rchestration; f. req. specification,'' 2016.

\bibitem{nfvper}
\BIBentryALTinterwordspacing
``{NFV}-{PER} 001.'' [Online]. Available:
  \url{http://www.etsi.org/deliver/etsi_gs/NFV-PER/001_099/001/01.01.01_60/gs_nfv-per001v010101p.pdf}
\BIBentrySTDinterwordspacing

\bibitem{kubernetes}
\BIBentryALTinterwordspacing
``Kubernetes.'' [Online]. Available: \url{http://kubernetes.io}
\BIBentrySTDinterwordspacing

\bibitem{xen}
\BIBentryALTinterwordspacing
``{Xen Project}.'' [Online]. Available: \url{http://www.xenproject.org}
\BIBentrySTDinterwordspacing

\bibitem{kvm}
\BIBentryALTinterwordspacing
``{KVM} {V}irtualization.'' [Online]. Available:
  \url{https://www.linux-kvm.org/page/Main_Page}
\BIBentrySTDinterwordspacing

\bibitem{stubdomains}
\BIBentryALTinterwordspacing
``Stub {D}omains.'' [Online]. Available:
  \url{http://www-archive.xenproject.org/files/xensummitboston08/SamThibault_XenSummit.pdf}
\BIBentrySTDinterwordspacing

\bibitem{osv}
\BIBentryALTinterwordspacing
``{OSv}.'' [Online]. Available: \url{http://osv.io}
\BIBentrySTDinterwordspacing

\bibitem{rumprun}
\BIBentryALTinterwordspacing
``{Rump Kernels}.'' [Online]. Available: \url{http://rumpkernel.org}
\BIBentrySTDinterwordspacing

\bibitem{scapy}
\BIBentryALTinterwordspacing
``Scapy.'' [Online]. Available: \url{https://scapy.net}
\BIBentrySTDinterwordspacing

\bibitem{osrally}
\BIBentryALTinterwordspacing
``{OpenStack Rally}.'' [Online]. Available:
  \url{https://wiki.openstack.org/wiki/Rally}
\BIBentrySTDinterwordspacing

\bibitem{vimdataset}
P.~Lungaroni \emph{et~al.}, ``{Results from Performance Evaluation and Testing
  of Virtual Infrastructure Managers},'' Zenodo,
  http://doi.org/10.5281/zenodo.1241097.

\bibitem{OpenStackCloudStack}
A.~Paradowski \emph{et~al.}, ``{Benchmarking the Performance of OpenStack and
  CloudStack},'' in \emph{2014 IEEE 17th International Symposium on
  Object/Component-Oriented Real-Time Distributed Computing}.\hskip 1em plus
  0.5em minus 0.4em\relax IEEE, 2014, pp. 405--412.

\bibitem{OpenStackvsEucalyptus}
D.~Steinmetz \emph{et~al.}, ``{Cloud Computing Performance Benchmarking and
  Virtual Machine Launch Time},'' SIGITE’12, pp. 89--90, 2012.

\bibitem{OpenStackOpenNebula}
E.~Caron \emph{et~al.}, ``{Comparison on OpenStack and OpenNebula performance
  to improve multi-Cloud architecture on cosmological simulation use case},''
  in \emph{Research Report RR-8421}.\hskip 1em plus 0.5em minus 0.4em\relax
  INRIA, 2013.

\bibitem{callegati2014performance}
G.~Callegati \emph{et~al.}, ``{Performance of Network Virtualization in cloud
  computing infrastructures: The OpenStack case},'' in \emph{IEEE CloudNet
  2014}, 2014.

\bibitem{litvinski2013experimental}
O.~Litvinski \emph{et~al.}, ``{Experimental evaluation of OpenStack compute
  scheduler},'' \emph{Proc. Computer Science}, vol.~19, pp. 116--123, 2013.

\bibitem{ledyayev2014high}
R.~Ledyayev \emph{et~al.}, ``{High performance computing in a cloud using
  OpenStack},'' \emph{CLOUD COMPUTING}, pp. 108--113, 2014.

\bibitem{cherrueau2017toward}
R.~Cherrueau \emph{et~al.}, ``{Toward a Holistic Framework for Conducting
  Scientific Evaluations of OpenStack},'' in \emph{17th IEEE/ACM International
  Symposium on Cluster, Cloud and Grid Computing}.\hskip 1em plus 0.5em minus
  0.4em\relax IEEE Press, 2017, pp. 544--548.

\bibitem{ieeeComms16rm}
R.~Mijumbi \emph{et~al.}, ``{Management and Orchestration Challenges in Network
  Function Virtualization},'' \emph{IEEE Communications Magazine}, vol. 54(1),
  2016.

\bibitem{ieeeNFVsurvey}
R.~Mijumbi \emph{et~al.}, ``{Network Function Virtualization: State-of-the-art
  and Research Challenges},'' \emph{IEEE Communications Surveys \& Tutorials},
  vol. 18(1), 2015.

\bibitem{tenor}
J.~Riera \emph{et~al.}, ``{TeNOR: Steps towards an orchestration platform for
  multi-PoP NFV deployment},'' in \emph{NetSoft Conference and Workshops
  (NetSoft), 2016 IEEE}.\hskip 1em plus 0.5em minus 0.4em\relax IEEE, 2016, pp.
  243--250.

\end{thebibliography}








\vspace{-5em}

\begin{IEEEbiography}[{\includegraphics[width=1in,height=1.25in,clip,keepaspectratio]{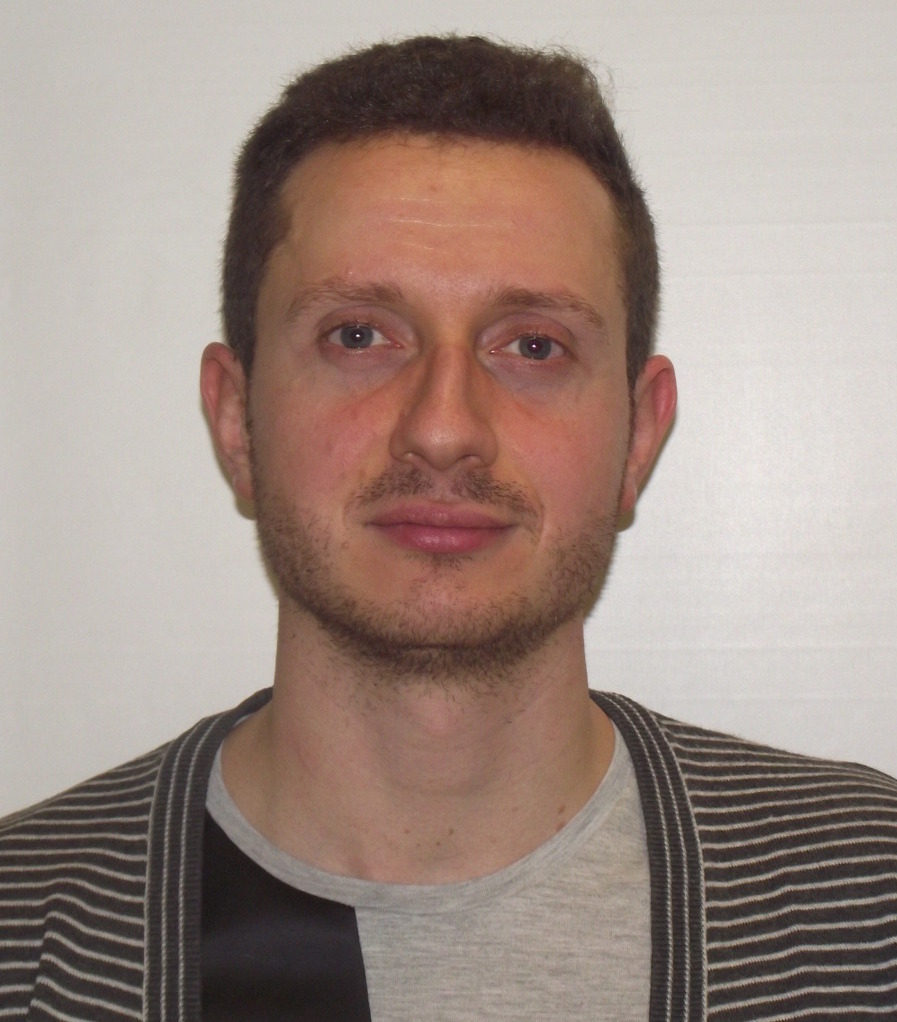}}]{Pier Luigi Ventre}
received his PhD in Electronics Engineering in 2018 from University of Rome ``Tor Vergata''. From 2013 to 2015, he was one of the beneficiary of the scholarship ``Orio Carlini'' granted by the Italian NREN GARR. His main research interests focus on Computer Networks, Software Defined Networking, Virtualization and Information-Centric Networking. He worked as researcher in several projects founded by the EU and currently he is a post-doctoral researcher at CNIT.
\end{IEEEbiography}

\vspace{-4em}

\begin{IEEEbiography}[{\includegraphics[width=1in,height=1.25in,clip,keepaspectratio]{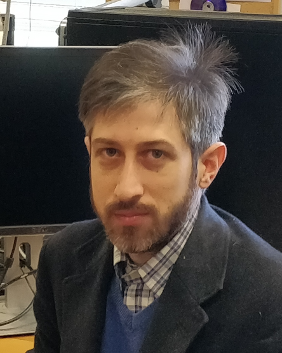}}]{Paolo Lungaroni}
received his Master's degree in Telecommunications Engineering from University of Rome ``Tor Vergata'' in 2015, with a thesis on Software Defined Networking Applications for the Network Function Virtualization Scenarios. From 2015 to 2018, he wins the scholarship ``Orio Carlini'' granted by the Italian NREN GARR. His main research interests focus on Computer Networks, Software Defined Networking and Virtualization. Currently, he works as researcher for CNIT.
\end{IEEEbiography}

\vspace{-4em}

\begin{IEEEbiography}[{\includegraphics[width=1in,height=1.25in,clip,keepaspectratio]{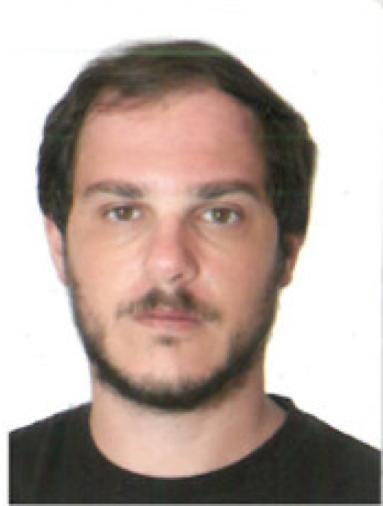}}]{Giuseppe Siracusano} 
is a researcher at NEC Laboratories Europe. He received his Master’s Degree in Computer Engineering from University of Rome ``Tor Vergata'' in 2013. He worked in different research and development projects as consultant for “Consorzio Nazionale Interuniversitario per le Telecomunicazioni”
(CNIT), currently he is also a PhD student in Electronic Engineering at University of Rome Tor Vergata. His main research interest are focused  software-based networking architectures and cloud networking.
\end{IEEEbiography}

\vspace{-4em}

\begin{IEEEbiography}[{\includegraphics[width=1in,height=1.25in,clip,keepaspectratio]{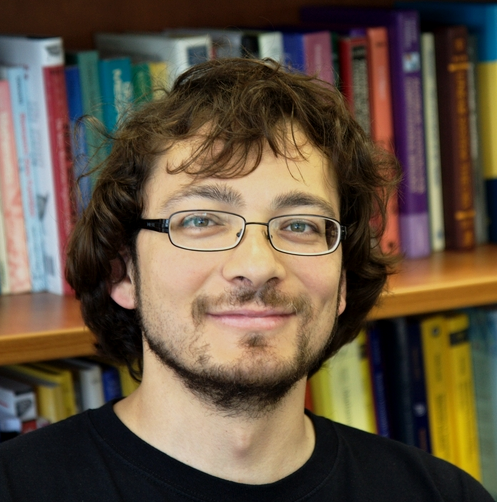}}]{Claudio Pisa}
received the M.Sc degree in Computer Science Engineering from the University ``Roma Tre'' in 2008, with a thesis on trusted routing in community networks. Then, after a short research collaboration, became a student at the University of Rome ``Tor Vergata'', receiving the Ph.D degree in 2013, with a thesis on Wireless Community Networks. Before joining CNIT as a researcher, he has worked as an R\&D engineer in the industry, on Software-Defined Networking (SDN), Cloud Computing and IoT projects.
\end{IEEEbiography}

\vspace{-4em}

\begin{IEEEbiography}[{\includegraphics[width=1in,height=1.25in,clip,keepaspectratio]{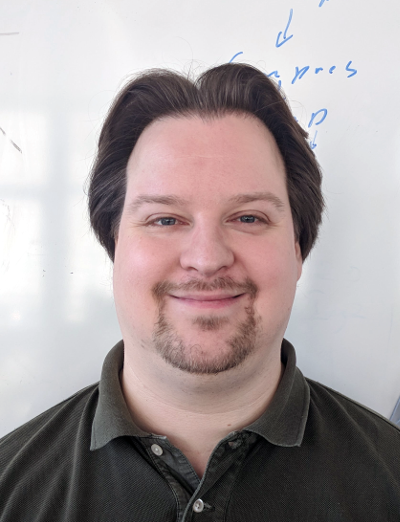}}]{Florian Schmidt}
received his PhD from RWTH Aachen in 2015 on the topic of heuristic error tolerance and recovery for network protocols.
His research background covers network protocol engineering, operating systems and virtualization, and wireless communications.
He currently is a researcher at NEC Laboratories Europe, where he focuses on research at the intersection of operating and networked systems and machine learning.
\end{IEEEbiography}

\vspace{-4em}

\begin{IEEEbiography}[{\includegraphics[width=1in,height=1.25in,clip,keepaspectratio]{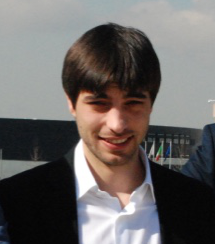}}]{Francesco Lombardo}
is co-founder and CTO at EveryUP S.r.l., spin off company of University of Rome ``Tor Vergata''. He received his Master's degree in Computer Engineering from University of Rome Tor Vergata in 2014. He has worked in different research and development projects as researcher for CNIT. His research interests include Network Function Virtualization, Software Defined Networking and mobile computing.
\end{IEEEbiography}

\vspace{-4em}

\begin{IEEEbiography}[{\includegraphics[width=1in,height=1.25in,clip,keepaspectratio]{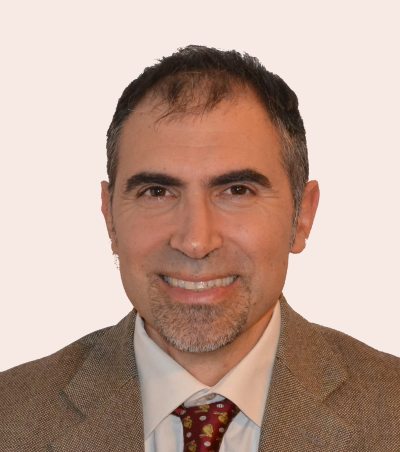}}]{Stefano Salsano}
(M'98-SM'13) received his PhD from University of Rome ``La Sapienza'' in 1998. He is Associate Professor at the University of Rome ``Tor Vergata''. He participated in 15 research projects founded by the EU, being project coordinator in one of them and technical coordinator in two. He has been PI in several research and technology transfer contracts funded by industries. 
His current research interests include SDN, Network Virtualization, Cybersecurity. He is co-author of an IETF RFC and of more than 150 peer-reviewed papers and book chapters.
\end{IEEEbiography}

\end{document}